\documentclass[twocolumn]{aastex631}

\newcommand{\kmps}{\, {\rm km \, s^{-1}}}

\begin{document}

\title{ALMA FACTS III. High-Resolution CO(2-1)/CO(1-0) Maps of Twelve Nearby Galaxies}

\author[0000-0001-8254-6768]{Amanda M Lee}
\affiliation{Stony Brook University, Stony Brook, NY 11743-3800, USA}
\author[0000-0002-8762-7863]{Jin Koda}
\affiliation{Stony Brook University, Stony Brook, NY 11743-3800, USA}

\author[0000-0002-1639-1515]{Fumi Egusa}
\affiliation{Institute of Astronomy, Graduate School of Science, The University of Tokyo, 2-21-1 Osawa, Mitaka, Tokyo 181-0015, Japan}

\author[0000-0002-0465-5421]{Akihiko Hirota}
\affiliation{National Astronomical Observatory of Japan, Los Abedules 3085 Office 701, Vitacura, Santiago 763 0414, Chile}
\affiliation{Joint ALMA Observatory, Alonso de C\'ordova 3107, Vitacura, Santiago 763 0355, Chile}

\author[0009-0007-2493-0973]{Shinya Komugi}
\affiliation{Division of Liberal Arts, Kogakuin University, 2665-1 Nakano-cho, Hachioji, Tokyo 192-0015, Japan}

\author[0000-0002-8868-1255]{Fumiya Maeda}
\affiliation{Research Center for Physics and Mathematics, Osaka Electro-Communication University, 18-8 Hatsucho, Neyagawa, Osaka, 572-8530, Japan}

\author[0000-0002-0588-5595]{Tsuyoshi Sawada}
\affiliation{National Astronomical Observatory of Japan, Los Abedules 3085 Office 701, Vitacura, Santiago 763 0414, Chile}
\affiliation{Joint ALMA Observatory, Alonso de C\'ordova 3107, Vitacura, Santiago 763 0355, Chile}

\shorttitle{CO(2-1)/CO(1-0) Variations in Twelve Galaxies}
\shortauthors{Lee et al.}

\begin{abstract}

We present early results from a high-resolution analysis ($\sim$100-200pc) of the CO(2-1)/CO(1-0) line ratio in twelve nearby galaxies. 
We use new ALMA CO(1-0) observations from the Fundamental CO(1-0) Transition Survey (FACTS), and re-imaged CO(2-1) data from PHANGS. 
We make empirical classifications based on the optical and molecular gas morphologies, which show clear systematic trends in the variation of $R_{21}$ as a function of galactic structure. 
The sample includes barred and unbarred, and flocculent galaxies.
The barred spiral galaxies follow a general trend when the gas exists significantly: $R_{21}$ is high in the center, low along the bar, increases at the bar ends, and then lowers beyond the bar end or flattens in the outer parts of the disk.
The structure dependence suggests the importance of galactic dynamics on molecular gas evolution, and consequently on star formation, in galaxies.
$R_{21}$ fluctuates in the spiral arms for both barred and unbarred galaxies. 
Areas around HII regions in some cases appear to show more high-ratio gas.
Together, $R_{21}$ varies systematically as a function of galactic structure, dynamics, and star formation activity.
\end{abstract}

\section{Introduction} \label{sec:intro}

The two lowest transitions, $J$=1-0 and 2-1, of $^{12}$CO are often used to trace the bulk molecular gas in galaxies. 
Both emissions likely originate from the same region within the molecular gas even when it is not resolved \citep[e.g.,][]{Sakamoto:1994, Sakamoto:1997, Oka:1998b}. Therefore, their line ratio CO(J=2-1)/CO(J=1-0) ($\equiv R_{21}$) can trace and constrain the physical conditions of the molecular gas \citep{Goldreich:1974, Scoville:1974, vanderTak:2007}. 
CO is excited by collisions with H$_{2}$ molecules, and its emissions travel through regions of high CO opacity.
Hence, $R_{21}$ is determined by the kinetic temperature $T_{\rm k}$, H$_{2}$ volume density $n_{\rm H_2}$, and the CO column density $N_{\rm CO}$. 
It can increase due to gas compression before star formation or due to gas heating after star formation \citep{Koda:2012}.
The variations of $R_{21}$ across galaxies can thus show the evolution of the bulk molecular gas in the galactic disks.

Studies in the Milky Way (MW) have shown systematic variations in $R_{21}$ within and amongst molecular clouds, and in the Galactic disk \citep[{see}][]{Hasegawa:1997}. 
These studies have classified molecular gas according to $R_{21}$ as low-ratio gas (LRG) with $R_{21} < 0.7$, high-ratio gas (HRG) with $R_{21} > 0.7$, and very-high ratio gas (VHRG) $R_{21} > 1.0$. 
For example, $R_{21}$ varies between and within star-forming and non-star-forming molecular clouds. 
\citet{Sakamoto:1994} and more recently \citet{Nishimura:2015} mapped the star-forming Orion molecular clouds, and found that $R_{21}$ ranges from 0.5 to 1.0 between the main ridge and peripheries of Orion A. 
\cite{Falgarone:1998} studied non-star-forming molecular clouds and found they have roughly uniform $R_{21}$ $\sim$ 0.65. 
In the Galactic disk, $R_{21}$ radially decreases \citep{Sakamoto:1997}, with azimuthal variations, i.e., LRG in the interarms and HRG in the spiral arms. 
In the Galactic center, the molecular gas exhibits distinctive properties than gas in the disk, and can be excited to high $R_{21}>0.7$, and occasionally, very-high ratios $R_{21}>1.0$ \citep{Oka:1996, Oka:1998b, Sawada:2001}. 
Overall, the ratio varies from 0.96 in the central kiloparsec region of the MW, while the typical disk value is 0.6-0.7 \citep{Sawada:2001}. 
VHRG with $R_{21} > 1.0$ is relatively rare and is typically found in the close vicinity of star formation, e.g. due to interaction with supernovae, or in the galactic center \citep{Oka:1996, Seta:1998}.

There have been many efforts to characterize $R_{21}$ in external galaxies. 
However, previous measurements were often hindered by challenges in calibration, especially since $R_{21}$ intrinsically has a relatively small dynamic range \citep[{only a factor of $\sim$2-3 variation; }][]{Garcia-Burillo:1993, Crosthwaite:2002, Lundgren:2004, Crosthwaite:2007, Koda:2012, Yajima:2021, Leroy:2022}. 
The calibration issue often led to high $R_{21}>1$ over large regions of the galaxies, seemingly indicating widespread optically-thin CO gas, which is not very likely. 
Only HRG $\sim$ 0.9 in galactic centers has been a consistent and physically-plausible result \citep[e.g.,][]{Braine:1992, Leroy:2009}.
Spatial variations in $R_{21}$ in external galaxies began to become apparent with improvements in calibration and techniques \citep[e.g.,][]{Koda:2012}. 
In particular, with the advent of ALMA, studies of $R_{21}$ have become increasingly possible \citep{Koda:2020, Egusa:2022, Maeda:2022, den-Brok:2023}.
There have also been galaxy-wide studies of $R_{21}$ in normal galaxies, \citep[e.g.,][]{Saintonge:2017, Keenan:2024, Keenan:2025}, as well as in extreme systems such as starburst and (ultra)-luminous infrared galaxies \citep[e.g.,][]{Papadopoulos:2012, Montoya-Arroyave:2023}.

High-resolution studies of $R_{21}$($\sim$ 100 pc) remain relatively scarce and are typically case-studies, or due to limited coverage, focus on particular regions in galaxies \citep[][]{Druard:2014, Egusa:2022, Maeda:2022, den-Brok:2023, Koda:2025}. 
These studies have found $R_{21}$ values similar to those of Galactic studies as well as indications of $R_{21}$ trends with respect to galactic structures. 
Beyond case-studies, it is clear that a survey of $R_{21}$ in nearby galaxies at high-resolution and with well-calibrated data, is needed to understand the general evolution of molecular gas and their physical conditions. 
While there are existing surveys of $R_{21}$ variations \citep[e.g.,][]{Yajima:2021, Leroy:2022, den-Brok:2025},  they are at lower sensitivity and resolution, {or} the compared CO(1-0) and CO(2-1) data were often observed from different telescopes.
As a result, overall systematic trends were not clear.

In this paper, we analyze $R_{21}$ in twelve nearby galaxies using new high-resolution CO(1-0) data from the ALMA-FACTS survey and reimaged CO(2-1) data from the PHANGS-ALMA survey \citep{Leroy:2021}. 
We present early results, which already empirically show systematic trends as a function of galactic structure.
This is the third paper of the ALMA-FACTS Survey. 
$R_{21}$ on kpc-scales in the same sample of galaxies using the ALMA Total Power (TP) component has been studied \citep[ALMA-FACTS Paper II;][]{Komugi:2025}. 
We plan multi-wavelength comparisons for future studies.

\section{Observations and Data Reduction} 
\label{sec:obsredu}
\begin{figure*}[h]
    \centering
    \rotatebox{90}{
    \begin{minipage}{\textheight}
    \centering 
    \includegraphics[width=1\textwidth]
    {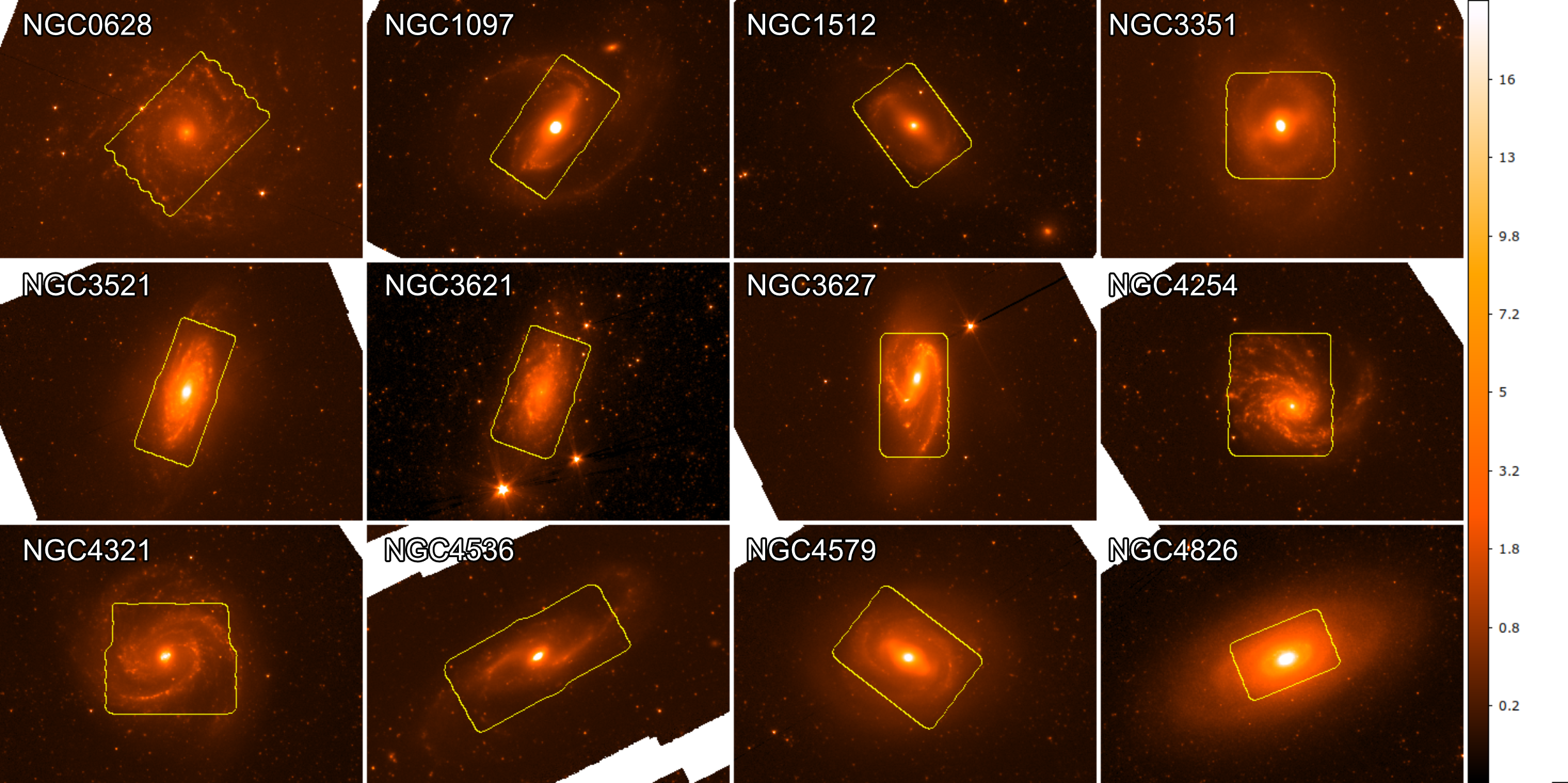}
    \caption{The sample shown in \emph{Spitzer} 3.6 $\mu$m emission [MJy/sr] from \citet[][]{Kennicutt:2003, Sheth:2010, S4G, SINGS} to demonstrate the field-of-view (FoV) of the 12m+07m CO(2-1) observations which are overlaid as yellow contours.}
    \label{fig:spitzer}
    \end{minipage}
    }
\end{figure*}

The FACTS sample of 12 galaxies (Table \ref{tab:sample}; Figure \ref{fig:spitzer}) is the complete set of all the galaxies common to the PHANGS-ALMA \citep{Leroy:2021}, Spitzer-SINGS \citep{Kennicutt:2003}, and Herschel-KINGFISH surveys \citep{Kennicutt:2011}, except one galaxy (NGC~4569) whose redshifted frequency is near the edge of the ALMA coverage and is not optimal for ALMA sensitivity.
It spans a wide range of spiral arm/bar strength and SF activity.
The design of the survey is discussed in a separate summary paper from the FACTS survey (Koda et al. in prep).

This section discusses the data reduction, mainly on the CO(2-1) data (Section \ref{sec:CO21red}).
The CO(1-0) data reduction is discussed in a separate paper (Koda et al. in prep.),
but is practically the same with one note in Section \ref{sec:CO10red}.
This method has been discussed in depth and used by \citet{Koda:2023} and \citet{Koda:2025}.

We used a common channel width of $5\kmps$ in CO(1-0) and CO(2-1).
After the imaging discussed below, we smoothed the CO(1-0) and CO(2-1) data to a common beam size of 2.5$\arcsec$ for the line ratio analysis.

Figure \ref{fig:spitzer} shows the sample in \emph{Spitzer} 3.6 $\mu$m dust emission from S4G \citep{Sheth:2010, S4G} and SINGS \citep[for NGC~3621,][]{Kennicutt:2003, SINGS} to demonstrate the field-of-view (FoV) of the observations. 
The FoVs of the 12m+07m CO(2-1) observations (yellow contours), and consequently the $R_{21}$ maps (section \ref{sec:r21}), cover the \emph{inner} disks of the galaxies. 

\subsection{CO(2-1)} \label{sec:CO21red}
The twelve galaxies in our sample were observed in CO(J=2-1) with the ALMA 12m, 07m, and TP arrays, as a part of the PHANGS survey \citep{Leroy:2021}. 
We obtained the raw archival data from projects 2012.1.00650.S (NGC~0628), 2015.1.00925.S (NGC~1512), 2015.1.00956.S (NGC~3351, 3627, 4254, and 4321), 2017.1.00886.L (NGC~1097, 3521, 3621, 4536, 4579, and 4826), and 2018.1.01651.S (NGC~1512). 
The procedure used for the TP data reduction is described in \citet{Koda:2020}. 
Here, we explain the data reduction of the 12m and 07m data.
The 12m and 07m data were calibrated with the Common Astronomy Software Application \citep[CASA;][]{CASA:2022} and imaged with Multichannel Image Reconstruction, Image Analysis, and Display \citep[MIRIAD;][]{Sault:1995, Sault:1996}.

\subsubsection{Calibration}
To calibrate the archival measurement sets (MSs), we ran the distributed pipeline scripts from the observatory, either generated for the PI, or a generic pipeline script. 
Different versions (v.4.7.2, v.5.1.2, and v.5.4.1) of the CASA pipeline were used across the galaxy sample. 
We used v.4.7.2 for projects 2012.1.00650.S and 2015.1.00925.S, v.4.7.2 and v.5.4.1 for 2015.1.00956.S, v.5.1.2 and v.5.4.1 for 2017.1.00886.L, and v.5.4.1 for 2018.1.01651.S. 

To assess the data after calibration, we checked the calibrated amplitudes and phases of the bandpass and gain calibrators (point sources). 
We considered the calibration successful if their amplitudes were flat and constant as functions of channel and time, and the phases as functions of channel and time were concentrated about 0 degrees. 
The consistency between the 12m and 07m calibrated data was checked by looking at the galaxy amplitude plotted against uv distance. 

For NGC~3621, the same gain calibrator was used across all of the archival 12m data (execution blocks) and had a weak amplitude of $\sim$60 mJy. 
This resulted in relatively large phase errors even after the calibration. 
In this case, we also made plots of the bandpass and gain calibrators in the complex plane and checked that their calibrated data points were localized. 

The standard flux calibration in the CASA pipeline was used for most galaxies.
However, NGC~0628 was observed relatively early in the ALMA observatory operation,
and the pipeline calibration returned too low a flux.
Therefore, we measured the fluxes of gain and bandpass calibrators with respect to Uranus or Neptune when they were observed in NGC~0628 measurement sets, and applied the measured fluxes to the rest of measurement sets when neither Uranus nor Neptune was included.

\subsubsection{Imaging}
The 12m and 07m array data for each galaxy were imaged together in MIRIAD, instead of CASA, because it takes into account spatial variations in the point spread function (PSF) \citep{Koda:2023}. 
To prepare for imaging in MIRIAD, we used the CASA tasks \emph{split} and \emph{mstransform} to split the narrowband spectral window (SPW) containing the CO(2-1) emission out from each 12m and 07m MS, and then to regrid them to a common velocity grid with channel width 5 km/s. 
These were converted into MIRIAD format. We used the average system temperature T$_{\rm sys}$ and integration times t$_{\rm int}$ from each MS following \citet{Koda:2023}. 
The invert task in MIRIAD was then performed on these 12m and 07m visibilities to produce the CO(2-1) data cubes. 
We used robust = +2 (approximately) corresponding to natural weighting), cellsize = 0.25’’, and a velocity coverage and width of $5\,\rm km/s$ matching those of the CO(1-0) data cubes (Koda et al. in prep).

We used the mossdi2 task in MIRIAD for cleaning. 
To identify regions where we may expect significant CO(2-1) emission, we first created masks for each galaxy by using its corresponding CO(1-0) signal-to-noise (SN) data cube (Koda et al. in prep). 
The mask includes connected regions with at least one pixel with SN $\geq$ 3.5$\sigma$ and 32 pixels with SN $\geq$ 2.5$\sigma$, and then is expanded to include pixels with SN $\geq$ 1.5$\sigma$. 
Using these masks, the dirty CO(2-1) cubes were cleaned in two steps: [1] within the mask and then [2] without a mask but cleaning the residual from [1] to recover emission that was potentially missed in the first step. 
We set a gain parameter of 0.05, 
options=positive (for cleaning within the mask in [1]), and 10$^7$ clean components per channel. 
The residual emission within the mask after the two-step cleaning were about $\sim$0.1$\sigma$. 

The cleaned 12m+07m data cubes were combined with the TP data cubes using the immerge task in MIRIAD. 
We identified velocity channels that were covered by both 12m+07m and TP data cubes, regridded the TP data spatially to the same grid as the 12m+07m data cube, and then combined the TP data cube to the primary-beam corrected 12m+07m cleaned data cube. 

The RMS noise per 5 km/s channel for the 12m+07m+TP data was measured with the dirty 12m+07m cube combined with TP data.
The RMS is measured in emission-free channels.
They are in Table \ref{tab:cubeparam}.

\subsection{CO(1-0)} \label{sec:CO10red}
The data reduction for CO(1-0) is discussed in a separate paper (Koda et al., in prep.), but is practically the same as the one discussed above.
The only difference is the very first step of imaging, where we did not have a reference cube to make a mask.
We went through the full imaging process without any mask and generated a 12m+07m+TP cube in CO(1-0).
A mask is then made with this cube for the final imaging which was done in the same way as discussed for CO(2-1).

\section{Maps} \label{sec:maps}

\begin{figure*}[h]
    \centering
    \rotatebox{90}{
    \begin{minipage}{\textheight}
    \centering
    \includegraphics[width=1\textwidth]
    {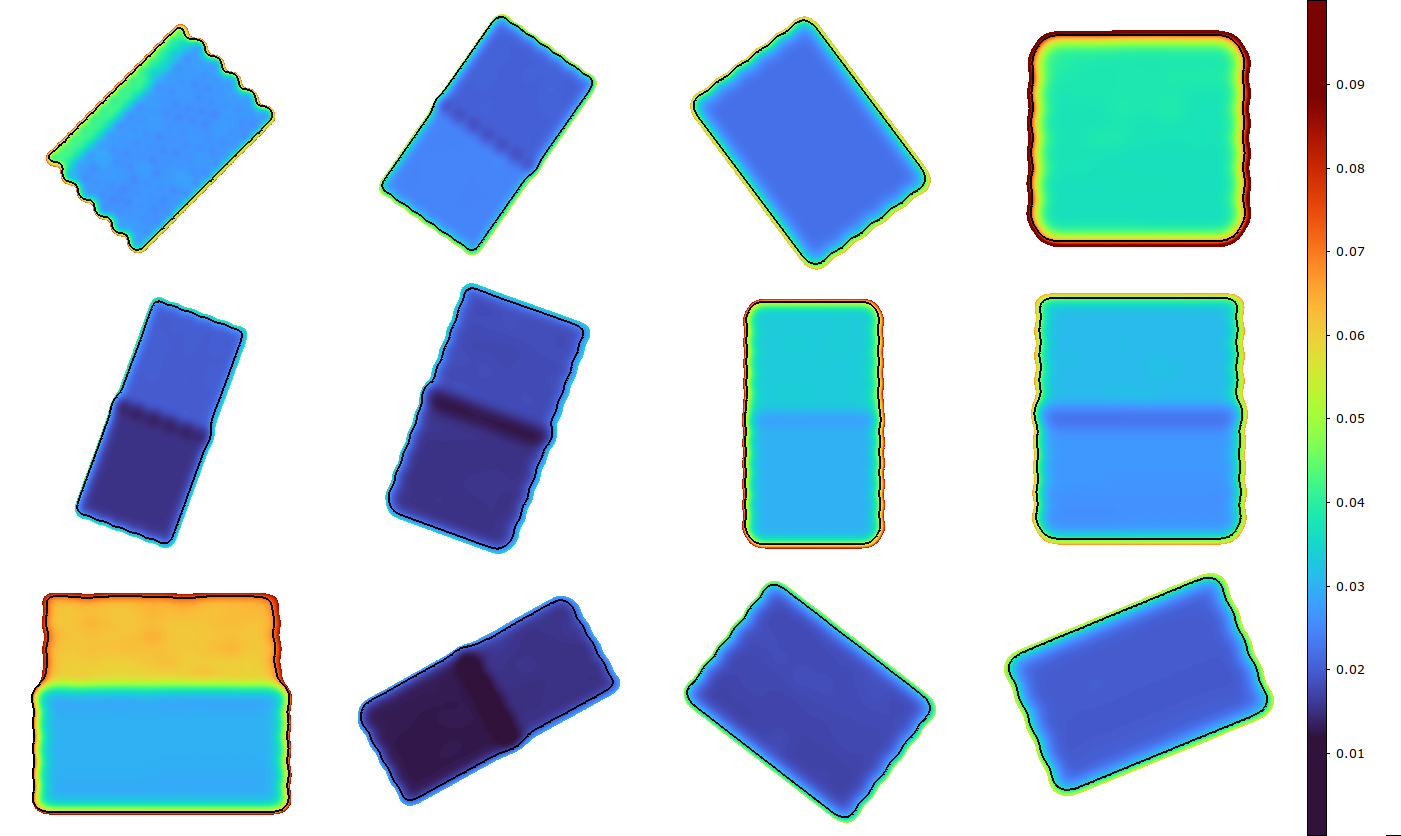}
    \caption{RMS [K] maps for 12m+07m+TP CO(2-1) (measured at the smoothed 2.5$\arcsec$ resolution) incorporating the sensitivity variations within each galaxy. 
    They are in the same arrangement as Figure \ref{fig:spitzer}. 
    The black contours correspond to when the sensitivities drops to 1/2 $\sim$ 50\% (40\% for NGC4321) of the peak sensitivity of each pointing.}
    \label{fig:senrm}
    \end{minipage}
    }
\end{figure*}

Figure \ref{fig:spitzer} shows the FoVs of the CO(2-1) data on the Spitzer $3.6\mu m$ images.
The corresponding RMS per channel maps for CO(2-1) are shown in Figure \ref{fig:senrm}. 
The CO(2-1) data were imaged out to when the sensitivity drops to $1/3 \sim 33\%$ of the peak sensitivity of each pointing (primary beam).
However, we restrict the analysis to the area where the sensitivity drops to  40-50\% of the peak sensitivity, which is indicated by the black contours in Figure \ref{fig:senrm} (yellow contours in Figure \ref{fig:spitzer}).
For most of the sample, multiple separate observations in CO(2-1) were taken to cover the entire field. 
Consequently, there are variations in the CO(2-1) sensitivity across the map for each galaxy, but for eleven out of the twelve galaxies, the sensitivities are roughly uniform within $\sim 20-30 \%$ across the entire field, and drop towards the edge. 
NGC~4321 is an exception with the northern part of the map having about twice lower sensitivity (i.e., twice higher noise) relative to the southern part.
NGC~4321 had only two 12m MSs, one for the northern part and the other covering the southern part.
Overall, to avoid the low sensitivity edges in the $R_{21}$ analysis, we only include the region where the sensitivity drops to $1/2 \sim 50\%$ of the peak sensitivity for eleven of the galaxies, and to $\sim 40\%$ of the peak sensitivity for NGC~4321.

\begin{figure*}[h]
    \centering
    \includegraphics[width=1\textwidth]    {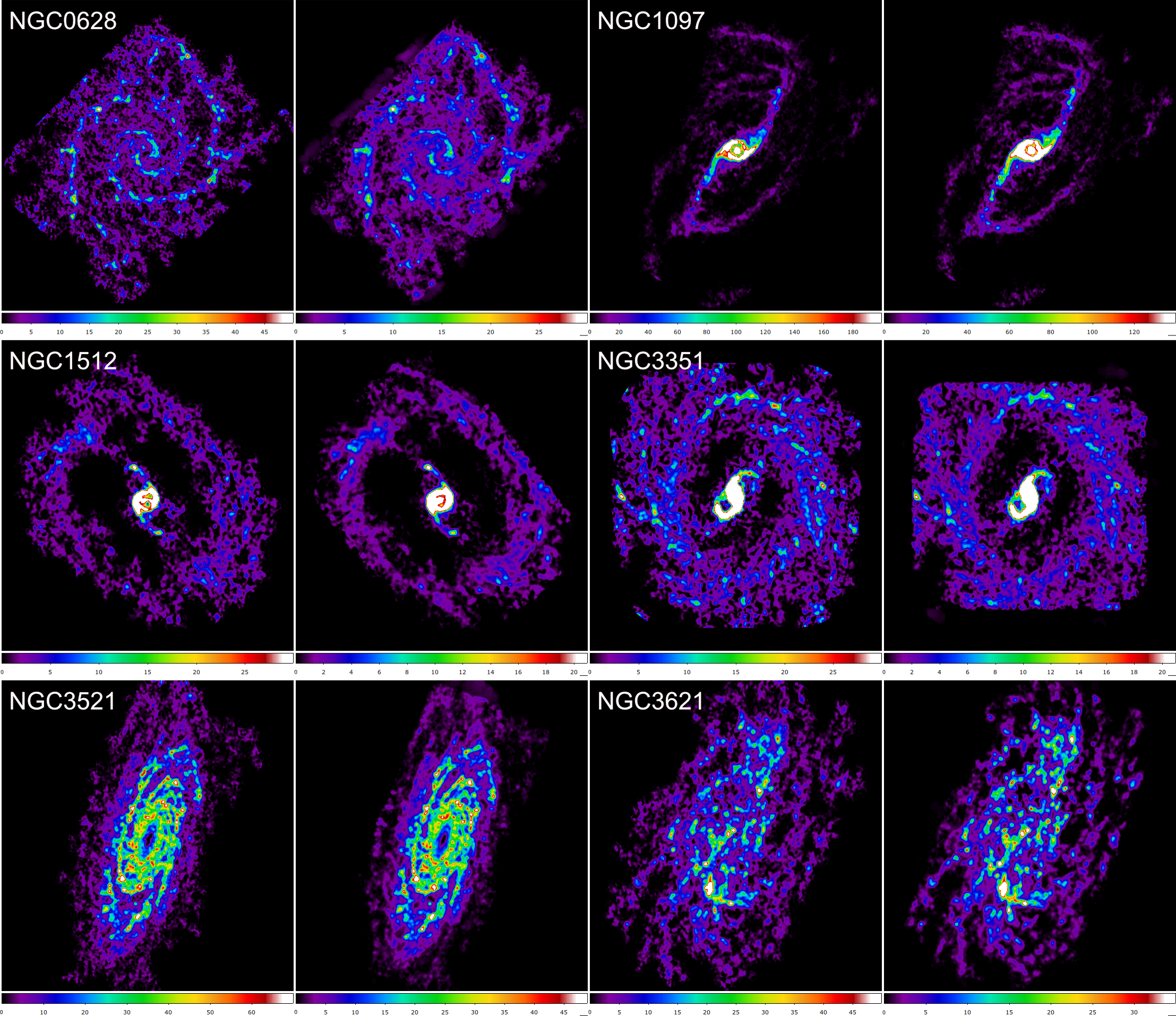}
    \caption{CO(1-0) (left) and CO(2-1) (right) integrated intensity maps [K$\cdot$km$\cdot$s$^{-1}$] both at 2.5$\arcsec$ resolution for NGC0628, NGC1097, NGC1512, NGC3351, NGC3521, and NGC3621 as labeled. The colorbar scale for CO(2-1) is adjusted so that the range covers 0.7 times that shown for CO(1-0).}
    \label{fig:co10co21_set1}
\end{figure*}

\begin{figure*}[h]
    \centering
    \includegraphics[width=1\textwidth]
    {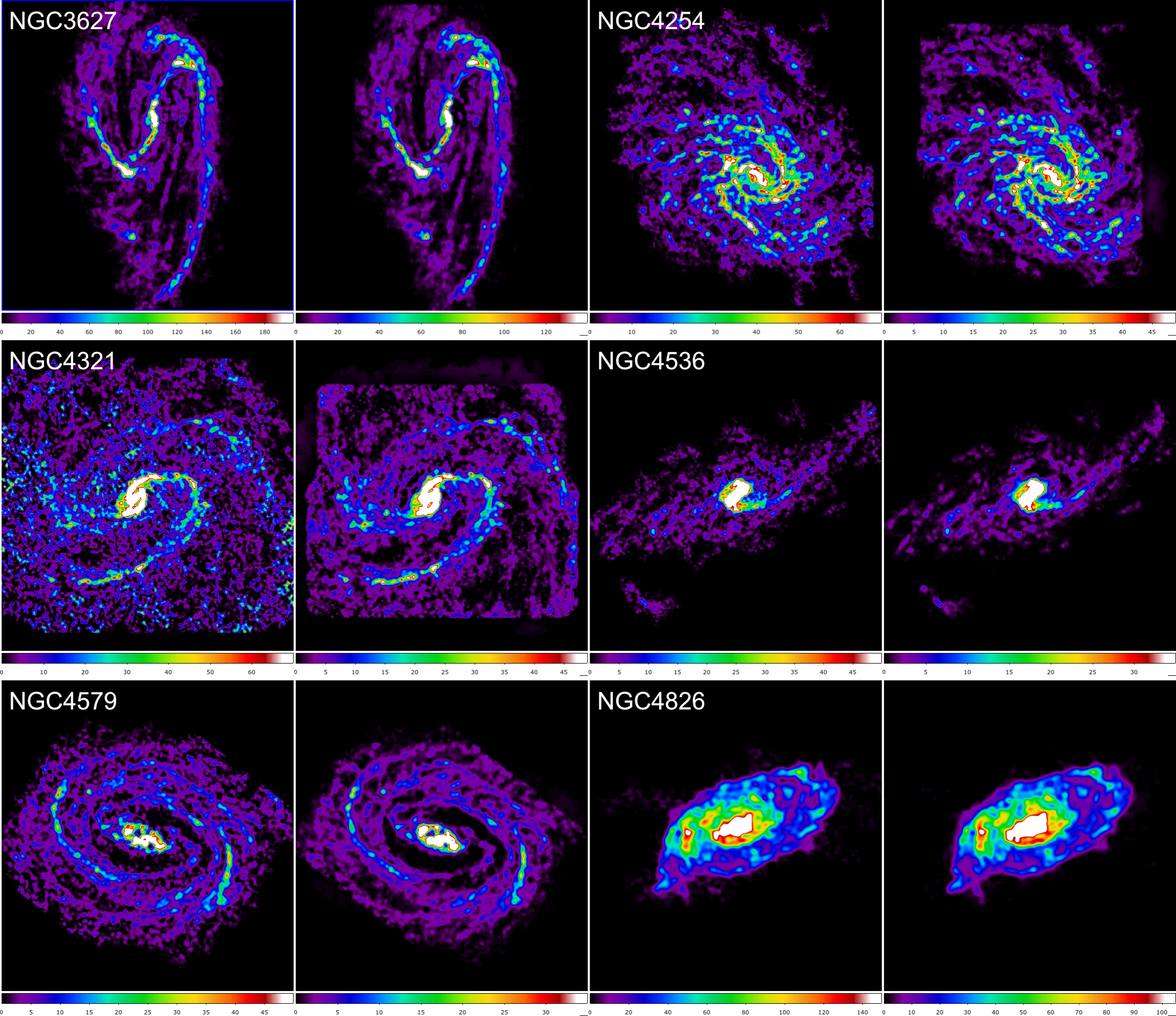}
    \caption{Same as Figure \ref{fig:co10co21_set1} but for NGC3627, NGC4254, NGC4321, NGC4536, NGC4579, NGC4826.}
    \label{fig:co10co21_set2}
\end{figure*}

Figures \ref{fig:co10co21_set1} and \ref{fig:co10co21_set2} show the integrated intensity maps, $I_{\rm{CO(1-0)}}$ and $I_{\rm{CO(2-1)}}$, in CO(1-0) and CO(2-1) for all the 12 galaxies.
The coverage of the CO(1-0) observations are slightly larger than the CO(2-1) observations.
Unlike CO(2-1), the CO(1-0) observations cover the entire field in each execution, and thus have more or less uniform sensitivity across the field. 
Hence, the resulting coverage and variations in sensitivity across the $R_{21}$ maps for each galaxy are mainly determined by the CO(2-1) observations.

To make the integrated intensity maps for each galaxy, we smoothed the CO(1-0) and CO(2-1) cubes to a common angular resolution of 2.5$\arcsec$.
We then regridded the CO(2-1) data spatially to the CO(1-0) data and applied a mask to the data cubes.
The mask was based on the (cleaned) CO(1-0) signal-to-noise (SN) cube starting from volumes that have at least 80 pixels with $>$ 5.0$\sigma$.
This mask was then expanded spatially by 2.5$\arcsec$ (=10 pixels in length) to include potential diffuse extended emission (if it exists).
The mask was not expanded in velocity.
After applying this mask, the cubes were integrated along the velocity axis to produce the integrated intensity maps, and corresponding error maps were generated.
The error in each pixel is calculated as RMS$\cdot\Delta v\cdot\sqrt{N}$, where RMS is the RMS noise per channel width $\Delta v$, and $N$ is the number of channels included in the velocity integration.

These are shown in Figures \ref{fig:co10co21_set1} and \ref{fig:co10co21_set2}.
The colorbar scales for $I_{\rm{CO(2-1)}}$ are adjusted to cover 0.7 times the shown range of $I_{\rm{CO(1-0)}}$ for a fairer comparison using a (fiducial) average line ratio $I_{\rm{CO(2-1)}}/I_{\rm{CO(1-0)}}$ of 0.7.
The $I_{\rm CO(1-0)}$ and $I_{\rm{CO(2-1)}}$ maps appear similar for prominent structures, although their ratio maps show variations (see Section \ref{sec:results}).

\begin{deluxetable*}{lccccccccc}
\tablecaption{Sample properties of ALMA-FACTS and adopted parameters for the bar. The corresponding spatial resolution at 2.5$\arcsec$.}
\tablewidth{0pt}
\tablehead{
\colhead{Galaxy}  & Morphology & Distance & $D_{25}$ & Inclination & PA$_{\rm disk}$ & $R_{\rm bar}$ & PA$_{\rm bar}$ & $\epsilon_{\rm bar}$ &  Scale (2.5$\arcsec$)\\
&    & [Mpc] & ($^{\prime}$) & [$^{\circ}$] & [$^{\circ}$]   & [$\arcsec$] & [$^{\circ}$] & & [pc] 
}
\decimalcolnumbers
\startdata
NGC~\ 0628      & SAc        & 9.8 (1)  & 10.5 &  19.8 &  20  &  -- & -- & -- & 119 \\
NGC~\ 1097      & SBb        & 15.4 (2) & 9.3  & 54.8 &  133.9 & 95.1  & 141 & 0.65 & 187 \\
NGC~\ 1512      & SBa        & 12.6 (2) & 8.9  & 68.3 &  65.2 & 73.5  & 42 & 0.66   & 153 \\
NGC~\ 3351      & SBb        & 9.33 (3) & 7.4  & 54.6 &  10.7 & 53.6 & 112 & 0.46   & 113 \\
NGC~\ 3521      & SA$^*$     & 12.1 (2) & 11   &     60.0 &  162.8 & -- & -- & --  &147 \\
NGC~\ 3621      & SAd        & 6.55 (3) & 12.3 &     67.6 &  161.7 & -- & --  & --  &79 \\
NGC~\ 3627      & SABb       & 9.38 (3) & 9.1 & 67.5 &  168.1  & 59.1 & 160 & 0.76 &   114\\
NGC~\ 4254      & SAc        & 13.9 (4) & 5.4  &     20.1 &  66  & --  & -- & -- & 168\\ 
NGC~\ 4321      & SABbc      & 14.3 (2) & 7.4  & 24.0 &  158 & 54.6 & 108 & 0.59  &   173 \\
NGC~\ 4536      & SABbc      & 14.5 (2) & 7.6  & 73.1 &  120.7 & 39.8 & 77 & 0.50 &  176 \\
NGC~\ 4579      & SABb       & 16.4 (4) & 5.9  & 41.9 & 90.2 & 42.2 & 53 & 0.48 &   199 \\ 
NGC~\ 4826      & SAab       & 4.41 (5) & 10   &      64.0 &  114.0  & -- & -- & -- & 53 \\
\enddata
\tablecomments{Columns (2-6) are from \citet[Paper II]{Komugi:2025}. \\
(1)~ Galaxy. \\ 
(2)~ Galaxy morphology taken from RC3 \citep{deVaucouleurs:1991}.  $^{*}$ For NGC~3521, \cite{deVaucouleurs:1991} lists this as a SABbc galaxy, but is treated as an unbarred SA for reasons explained in \citet{Komugi:2025}. \\
(3)~ Redshift-independent distance. Numbers in parentheses is the reference.  In order of preference, measurements taken from Cepheid variables, tip of RGB, Tully-Fischer relation.  References : 1. \citet{McQuinn:2017}, 2. \citet{Tully:2016}, 3. \cite{Freedman:2001}, 4. \cite{Tully:2013}, 5. \citet{Anand:2021}. \\
(4)~ Isophotal diameter at the major axis, taken from RC3 \citep{deVaucouleurs:1991}. \\
(5)(6)~ Inclination and position angle taken from HyperLEDA \citep{Makarov:2014} when available, or from \cite{Kuno:2007}, \cite{Sorai:2019}. \\ 
(7-9)~ Adopted semi-major axis, position-angle, and ellipticity of the bar from \citet{Herrera-Endoqui:2015} for bar region masks (section \ref{sec:r21_structures}). \\ 
(10) Spatial resolution at 2.5$\arcsec$.} 
\label{tab:sample}
\end{deluxetable*}

\section{Results \label{sec:results}}

From the integrated intensity maps in Section \ref{sec:maps}, we calculated the line ratio $R_{21}\equiv{I_{\rm{CO(2-1)}}}/{I_{\rm{CO(1-0)}}}$ maps (Figures \ref{fig:r21rgb_set1}-\ref{fig:r21rgb_set6}). We applied a signal-to-noise (SN) cut of $>$5 in both CO(1-0) and CO(2-1).
We summarize the $R_{21}$ values and their distribution for each galaxy individually. 
The spatial resolutions of our maps at the adopted angular resolution of 2.5$\arcsec$ range from $\sim$ 50-200 pc amongst the sample (Table \ref{tab:sample}), larger than the typical size of a molecular cloud. 
The $R_{21}$ values presented here thus represent the average $R_{21}$ within individual molecular clouds.

\subsection{$R_{21}$ Maps and Distributions \label{sec:r21}}

The $R_{21}$ maps for each galaxy and their distributions (histograms), as well as their optical images for comparison, are shown in Figures \ref{fig:r21rgb_set1}-\ref{fig:r21rgb_set6}. 
We summarize the $R_{21}$ variations in each galaxy separately (see Table \ref{tab:percentages}).
We tend to find similar trends in each morphological type and galactic structure (e.g., barred, unbarred, spiral), which leads us to classify them according to their optical and molecular gas morphologies (Section \ref{sec:disc}).

We divide the molecular gas using the definitions adopted in \citet[]{Hasegawa:1997}. 
Low-ratio gas (LRG) have $R_{21} < 0.7$ (blue), high-ratio gas (HRG) are in between $0.7 < R_{21} < 1.0$ (red), and very-high-ratio gas (VHRG) have $R_{21} > 1.0$. 
The division of molecular gas at $R_{21}=0.7$ (white) is motivated both theoretically via the large-velocity gradient (LVG) approximation \citep{Goldreich:1974, Scoville:1974} and observationally first in MW studies \citep[e.g.,][]{Sakamoto:1997, Oka:1998a, Hasegawa:1997}.
Typical molecular gas with physical conditions (${n_{\rm{H_2}}, T_{\rm{k}}}$) = (300 cm$^{-3}$, 10 K) have $R_{21}$ $\sim$ 0.6-0.7 (around the LRG-HRG boundary), with the transition from LRG to HRG requiring 2-3x denser and/or hotter gas \citep[e.g.,][]{Koda:2012}. 
On the other hand, VHRG with $R_{21} > 1.0$ is difficult to achieve, requiring optically thin conditions yet even denser/hotter gas. 

We note that the shift from LRG to HRG (or vice-versa) is continuous in the values of $n_{\rm H2}$, $T_{\rm k}$ (and $N_{\rm CO}$) \citep{Koda:2025} and cannot be characterized by single numbers. 
Nevertheless, a crude guideline would be helpful to gauge the relative physical conditions of LRG and HRG.
According to the calculations by \citet{Koda:2012} which considered a range of $N_{\rm CO}$, HRG requires $\sim$2-3 times higher $n_{\rm H2}$ and $T_{\rm k}$ than LRG.
 
The blue and green histograms show how the bulk of the molecular gas in each galaxy are distributed by area (number) and mass (weighted by I$_{\rm{CO(1-0)}}$), respectively. 
They show the overall distribution (lightest shade) and the contribution from the central 1~kpc ($R<0.5$~kpc, darkest shade) in the plane of each galaxy, following the definition for central regions adopted in e.g. \citet[][]{Sakamoto:1999, Sheth:2005}.
The central 1~kpc has been used as a fiducial value for central regions in the literature. 
However, the size of the central region depends on the galaxy, and based on the $R_{21}$ maps, the region of HRG in the centers of galaxies can extend beyond $R>0.5$~kpc. 
We thus also show the contribution from the central 2~kpc ($R>1.0$~kpc, hatched, dark shade) in the histograms.

\subsubsection{NGC~0628}

NGC~0628 is an isolated spiral galaxy (SAc) with many star-forming regions (HII) along its spiral arms.
It does not have a bar structure.
At its distance, the angular resolution 2.5$\arcsec$ corresponds to a spatial resolution of 119 pc.
From Figure \ref{fig:co10co21_set1}, the molecular gas distribution in NGC~0628 is relatively dim (e.g. compared to the other unbarred galaxies in the sample) in both transitions of CO and its spiral arms appear loosely wound. 
The stellar spiral arms extend beyond the CO coverage (e.g., Figure \ref{fig:spitzer}). 

\begin{figure*}[h]
    \centering
    \includegraphics[width=1\textwidth]{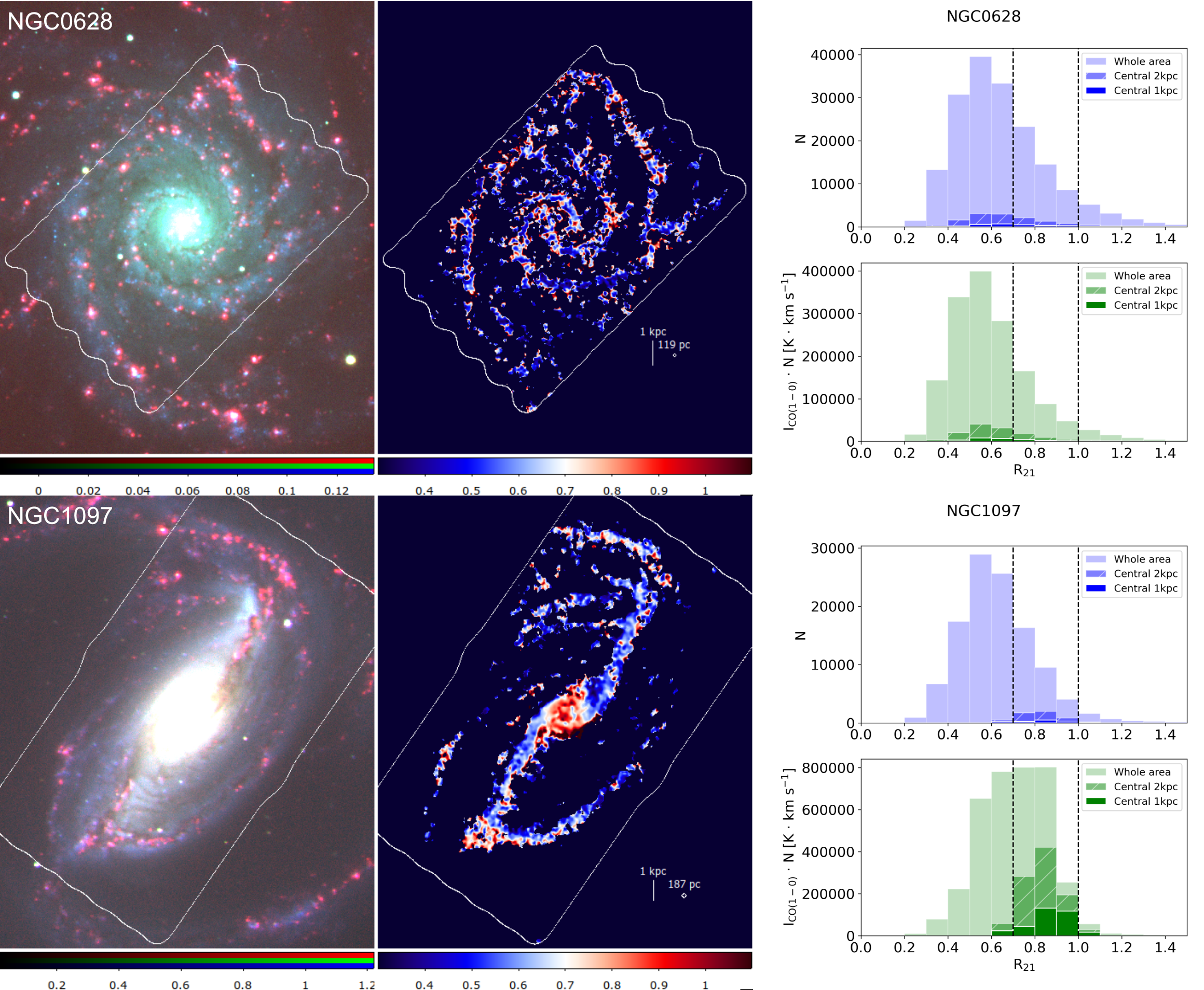}
    \caption{Optical (B, V, H$\alpha$) image and map of $R_{21}$ and its distribution by number (blue histograms) and weighted by CO(1-0) intensity (green histograms) for NGC0628 and NGC1097. 
    The histograms show the overall distributions (lightest shade) and contributions from the central 2~kpc ($R<1.0$~kpc, hatched, dark shade), and central 1~kpc ($R<0.5$~kpc, darkest shade) in the plane of the galaxy.
    The colormap follows the $R_{21}$ classification of molecular gas from \citet[][]{Hasegawa:1997}, with blue for $R_{21}$ $<$ 0.7 (low-ratio gas), white for $R_{21}$=0.7, and red for R$_{21}$ $>$ 0.7 (high-ratio gas). The optical images were obtained from SINGS \citep[][]{Kennicutt:2003, SINGS}. The white contours are the FoVs of the 12m+07m CO(2-1) observations. The black dashed lines in the histograms are at $R_{21}$=0.7 and 1.0 for reference. 
    The RGB image scale here is by eye.}
    \label{fig:r21rgb_set1}
\end{figure*}

\begin{figure*}[h]
    \centering
    \includegraphics[width=1\textwidth]
    {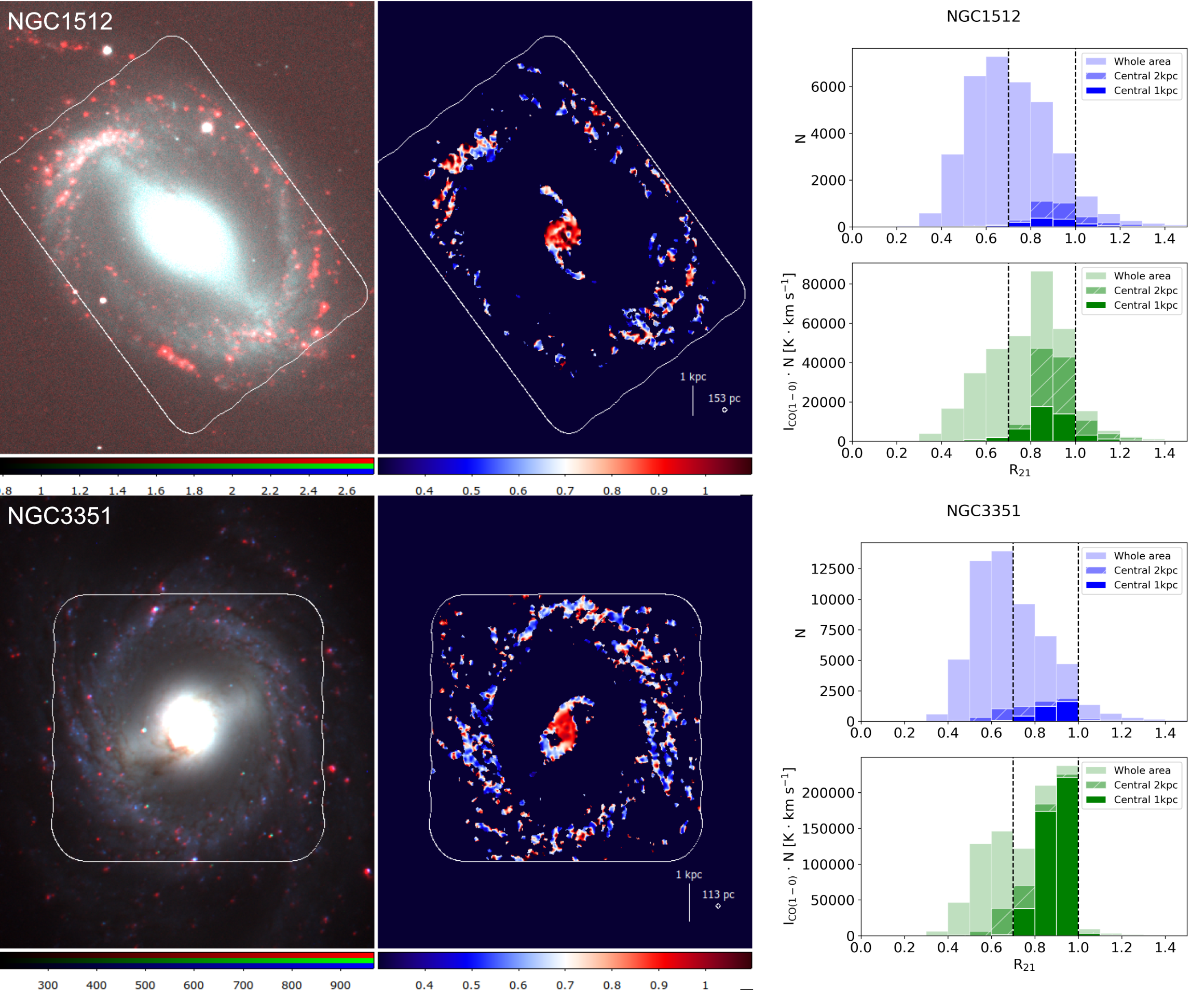}
    \caption{Same as Figure \ref{fig:r21rgb_set1} for NGC1512 and NGC3351.}
    \label{fig:r21rgb_set2}
\end{figure*}

\begin{figure*}[h]
    \centering
    \includegraphics[width=1\textwidth]
    {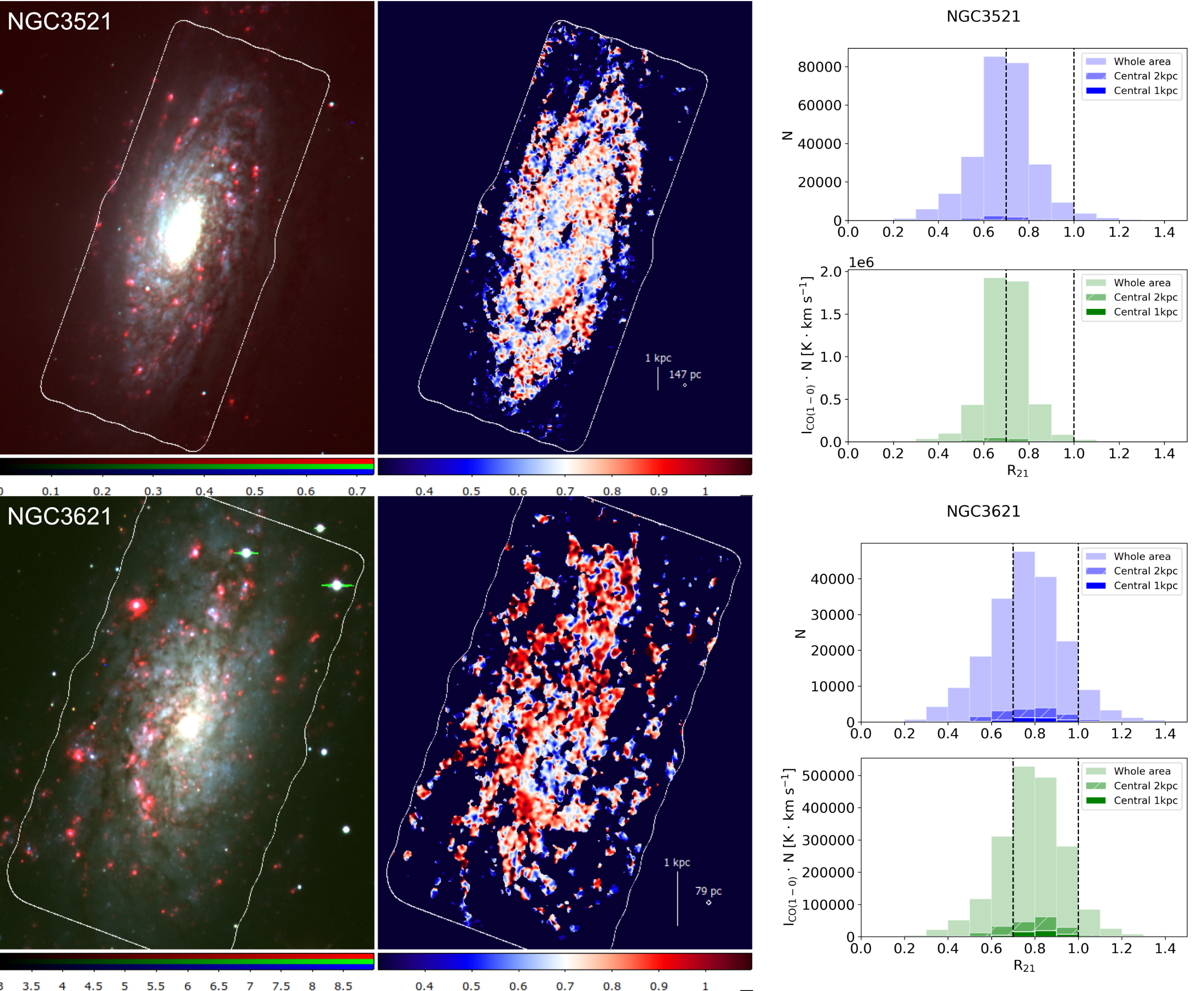}
    \caption{Same as Figure \ref{fig:r21rgb_set1} for NGC3521 and NGC3621.}
    \label{fig:r21rgb_set3}
\end{figure*}

\begin{figure*}[h]
    \centering
    \includegraphics[width=1\textwidth]
    {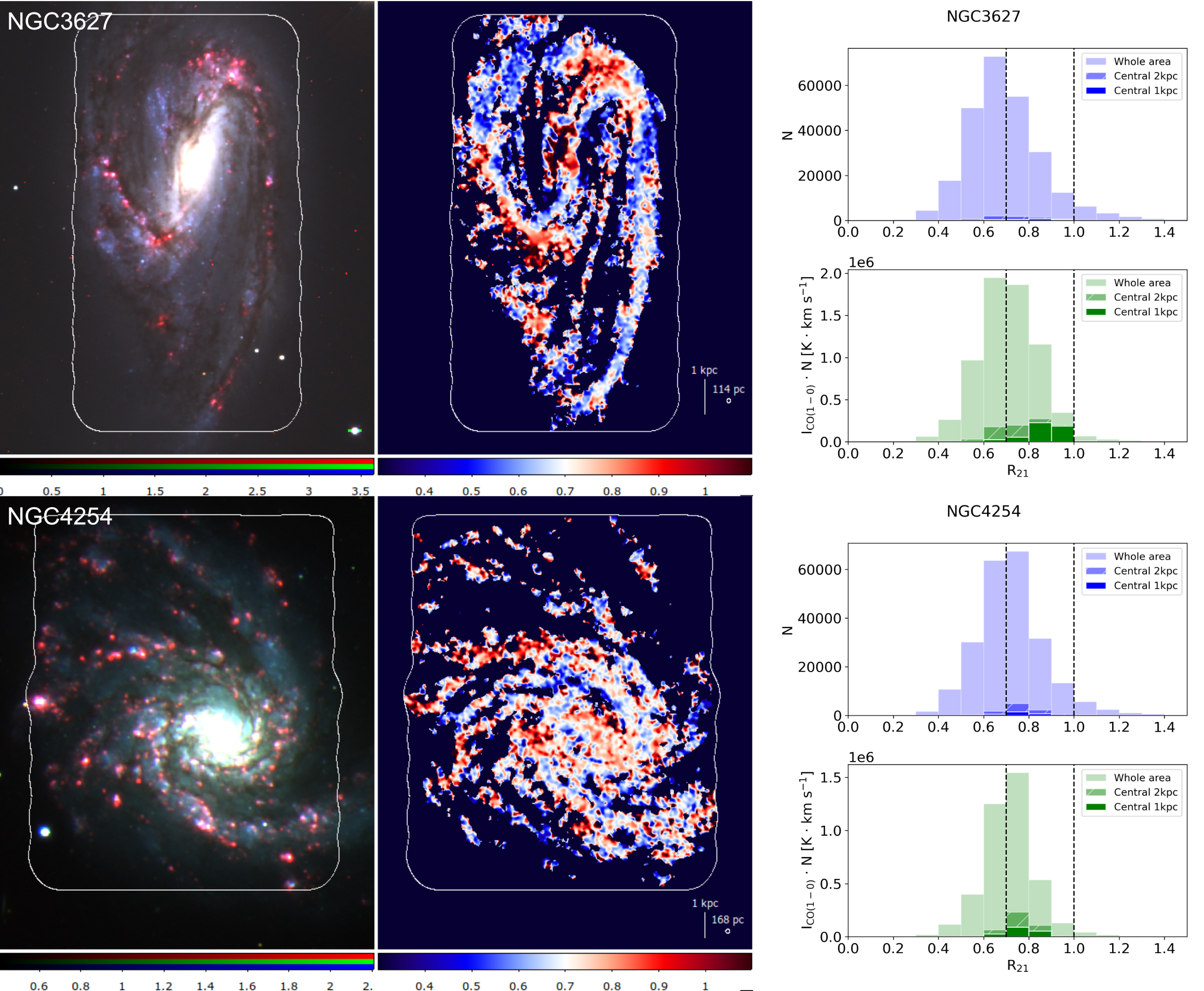}
    \caption{Same as Figure \ref{fig:r21rgb_set1} for NGC3627 and NGC4254.}
    \label{fig:r21rgb_set4}
\end{figure*}

\begin{figure*}[h]
    \centering
    \includegraphics[width=1\textwidth]
    {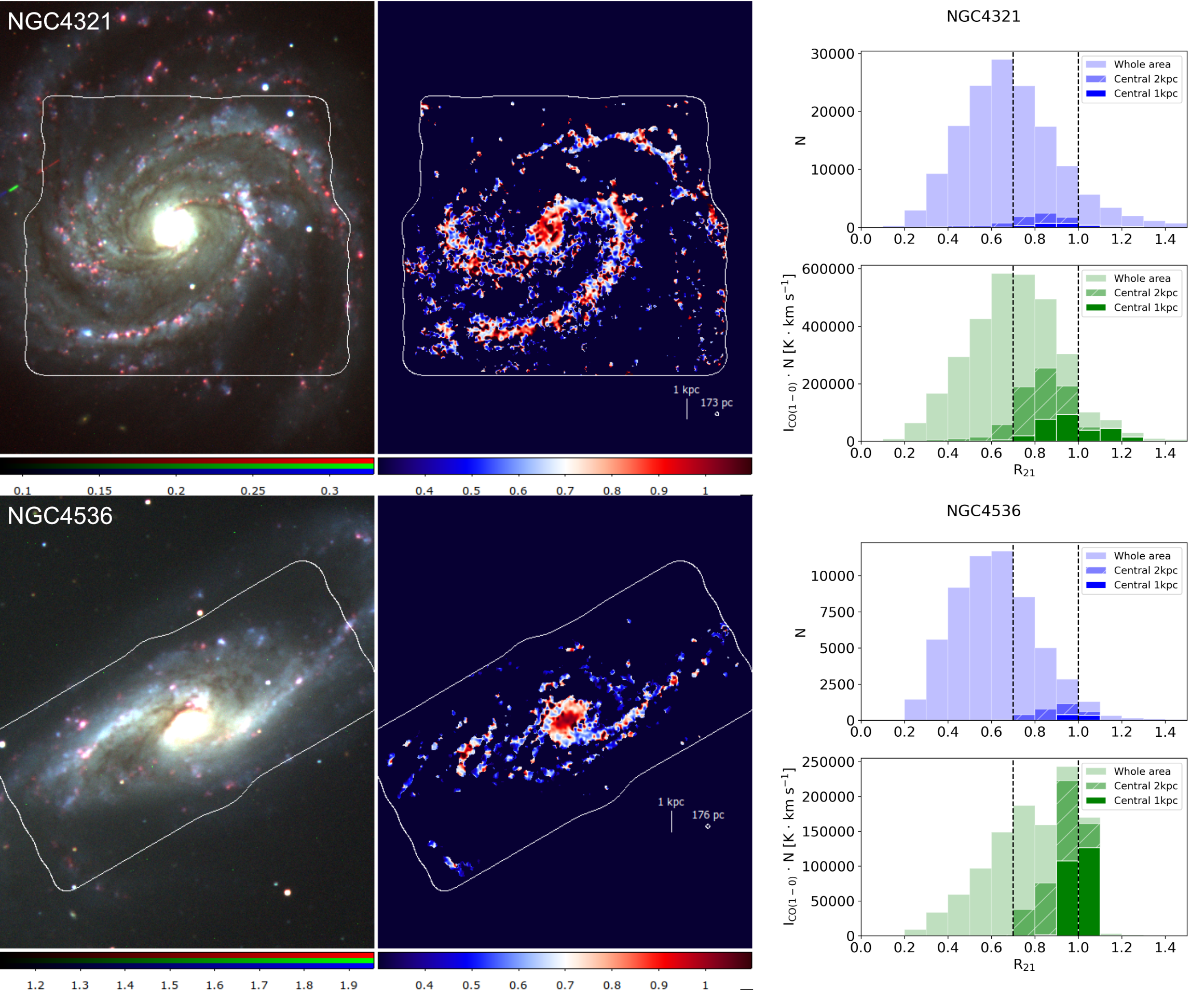}
    \caption{Same as Figure \ref{fig:r21rgb_set1} for NGC4321 and NGC4536.}
    \label{fig:r21rgb_set5}
\end{figure*}

\begin{figure*}[h]
    \centering
    \includegraphics[width=1\textwidth]
    {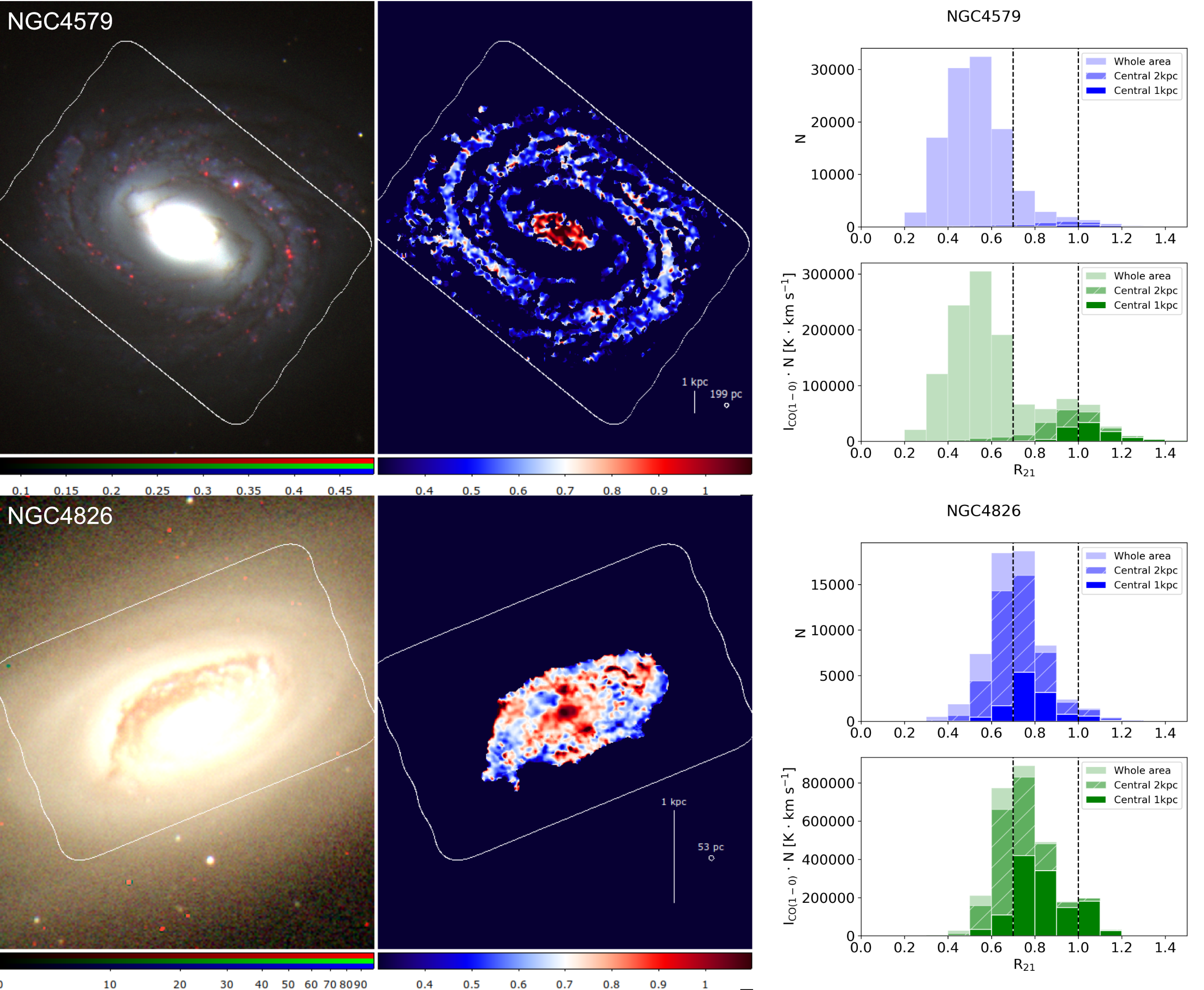}
    \caption{Same as Figure \ref{fig:r21rgb_set1} for NGC4579 and NGC4826.}
    \label{fig:r21rgb_set6}
\end{figure*}

The $R_{21}$ measurements are concentrated around the spiral arms due to the sensitivity limitation, with that in the interarm regions only occasionally determined.
The disk is dominated by low-ratio gas (Figure \ref{fig:r21rgb_set1}), encompassing 67\% of the significantly detected pixels, while high-ratio gas makes up 26\% (blue histogram). 
This is similar for the CO(1-0) intensity weighted distribution (green histogram), with low-ratio gas and high-ratio gas containing 77\% and 19\% respectively.
Over the whole area, the mean $R_{21}$ is 0.65 while the intensity-weighted mean is 0.60.

\citet{Barnes:2023} conducted a detailed case-study of the Phantom Void (1~kpc bubble) in this galaxy. 
They suggested the bubble is sustained by stellar feedback. 
In Figure \ref{fig:r21rgb_set1}, the molecular gas around there does not show a particular excess in $R_{21}$.

\subsubsection{NGC~1097}

NGC~1097 is a strongly barred, interacting spiral galaxy (SBb).
At its distance, the angular resolution of 2.5$\arcsec$ corresponds to a spatial resolution of 187 pc. 
It is the second-farthest galaxy in the sample. 
From Figure \ref{fig:co10co21_set1}, the molecular gas distribution shows a prominent central core and offset ridges along the bar. 
The spiral arms extend from the bar ends beyond our coverage (e.g., Figure \ref{fig:spitzer}). 
There are also interarm structures (structures between the offset ridges) that often appear filamentary. 

$R_{21}$ is high in the center, becomes lower along the bar, and then increases at the bar ends (Figure \ref{fig:r21rgb_set1}). 
Low-ratio gas and high-ratio gas make up 70\% and 26\% of the area (blue histogram), respectively. 
However, the CO(1-0) intensity contained by low-ratio and high-ratio gas are similar, at 47\% and 50\% respectively (green histogram). 
Over the whole area, the mean $R_{21}$ is 0.63 while the intensity-weighted mean is 0.71.

From the dark/hatched and darkest regions in the histograms, the central 1-2~kpc ($R<1$~kpc) contains mostly high-ratio gas.
The middle panel shows a 1-kpc scale (white bar) in the lower-right corner.
The high-ratio gas region, and the central gas concentration, in this galaxy extends even beyond the traditional definition of the central 1~kpc region \citep{Sakamoto:1999, Sheth:2005}.
In any case, the whole region is located at the roots of the offset ridges.

\subsubsection{NGC~1512}

NGC~1512 is a barred spiral galaxy (SBa) with a companion. 
At its distance, the angular resolution of 2.5$\arcsec$ corresponds to a spatial resolution of 153 pc. 
From Figure \ref{fig:co10co21_set1}, the molecular gas distribution in NGC~1512 shows a bright core and two offset ridges.
The ridges are significantly detected in their inner parts but not in the outer parts.
The CO emission re-appears around the bar ends, which continues to fainter spiral arms.

$R_{21}$ is high in the center and becomes lower along the offset ridges (Figure \ref{fig:r21rgb_set2}). 
The $R_{21}$ along the spiral arms fluctuate between low-ratio and high-ratio, with some high-ratio gas spatially coincident with HII regions. 
This galaxy also has a central gas concentration and high ratio region larger than the nominal definition of the central 1~kpc.
Indeed, the histogram for the central 2~kpc explains most of the HRG.
Low-ratio and high-ratio gas encompass 50\% and 42\% of the area shown, respectively.
However, low-ratio gas contain 31\% of the CO(1-0) intensity, while 61\% is contained within the high-ratio gas. 
Over the whole area, the mean $R_{21}$ is 0.72 while the intensity-weighted mean is 0.79.
From the histograms, the central 1-2~kpc ($R<1.0$ kpc) contain mostly high-ratio gas.

\subsubsection{NGC~3351}

NGC~3351 is a barred spiral galaxy (SBb). 
The angular resolution of 2.5$\arcsec$ corresponds to 113 pc. 
From Figure \ref{fig:co10co21_set1}, the molecular gas distribution in NGC~3351 has a bright central core and two offset ridges, also surrounded by spiral arms.
The offset ridges are bright in the inner part of the bar, but not in the outer part.

In Figure \ref{fig:r21rgb_set2}, $R_{21}$ is high in the center and decreases along the offset ridges, but fluctuates in the surrounding spiral arms between low and high ratios. 
Low-ratio and high-ratio gas make up 58\% and 38\% of the area shown (blue histogram). 
However, the amount of 
low-ratio gas and high-ratio gas contained by the CO(1-0) intensity are  36\% and 62\%, respectively (green histogram).
Over the whole area, the mean $R_{21}$ is 0.69 while the intensity-weighted mean is 0.77.
From the histograms, high-ratio gas is concentrated predominantly in the central kiloparsec ($R<0.5$ kpc) region. 

\citet{Teng:2022} studied $^{12}{\rm CO}$, $^{13}{\rm CO}$ and $^{18}{\rm CO}$ line ratios (at 2.1$\arcsec$ resolution; $\sim$ 100~pc) in the central kiloparsec of NGC~3351.
They also found that the central region consists of high-ratio gas $R_{21} \sim 1.0$.

\subsubsection{NGC~3521}

NGC~3521 is a spiral galaxy (SA) with star-forming (HII) regions spread across its disk. 
This is an unbarred galaxy.
The resolution of 2.5$\arcsec$ corresponds to 147 pc. 
The optical image of NGC~3521 shows a flocculent pattern across the disk, but from Figure \ref{fig:co10co21_set1}, the molecular gas distribution shows tightly wound spiral arms. 
Both transitions of CO are detected across the disk, but there is no central gas concentration as seen in the barred galaxies.

$R_{21}$ fluctuates between low and high values along the spiral arms (Figure \ref{fig:r21rgb_set3}). 
Low-ratio and high-ratio encompass 52\% and 45\% of the disk respectively (blue histogram). 
The CO(1-0) intensity contained by the low-ratio and high-ratio gas are 50\% and 49\% respectively (green histogram).
Over the whole area, the mean $R_{21}$ is 0.69 while the intensity-weighted mean is 0.70.
As both transitions of CO are not detected in the center, there is little contribution from the central kiloparsec ($R<0.5$~kpc) in the $R_{21}$ distributions, if any. 

\subsubsection{NGC~3621}

NGC~3621 is another isolated unbarred spiral galaxy (SAd) with star-forming regions spread across its disk. 
At its distance, the angular resolution of 2.5$\arcsec$ corresponds to a spatial resolution of 79 pc. 
It is the second-closest galaxy in the sample.

From Figure \ref{fig:co10co21_set1}, 
the molecular gas structure appears less organized and flocculent.
$R_{21}$ fluctuates between low and high values across the disk (Figure \ref{fig:r21rgb_set3}). 
The spatial distribution of high-ratio gas seems to coincide closely with HII regions. 
Low-ratio gas and high-ratio gas cover 35\% and 58\% of the area shown (blue histogram). 
The CO(1-0) intensity contained by low-ratio and high-ratio gas are 26\% and 67\% (green histogram). 
Hence, the majority of the molecular gas by area and by CO(1-0) intensity have high-ratios.
Over the whole area, the mean $R_{21}$ is 0.76 while the intensity-weighted mean is 0.78.

\subsubsection{NGC~3627}

NGC~3627 is an interacting barred spiral galaxy (SABb) in the Leo Triplet group. 
At its distance, the angular resolution of 2.5$\arcsec$ corresponds to a spatial resolution of 114 pc. 
From Figure \ref{fig:co10co21_set2}, the molecular gas distribution in NGC~3627 shows a prominent center and bar, with coherent spiral arms extending from the bar ends. 
The central region and bar ends are bright in both CO transitions. 

$R_{21}$ is high in the center, lowers along the bar, and then increases again at the bar ends (Figure \ref{fig:r21rgb_set4}). 
The high ratio regions especially at the bar ends and along the spiral arms are located near the HII regions in the optical image. 
Low-ratio and high-ratio gas encompass 57\% and 38\% of the galaxy by area (blue histogram), while they contain a similar amount of the CO(1-0) intensity at 48\% and 50\% (green histogram), respectively. 
Over the whole area, the mean $R_{21}$ is 0.70 while the intensity-weighted mean is 0.71.
The majority of the gas in the central kiloparsec ($R<0.5$ kpc) is high-ratio gas.

\citet[][]{den-Brok:2023} also studied $R_{21}$ in NGC~3627 (as well as other line ratios including CO(3-2) ALMA observations) at $4\arcsec$ resolution ($\sim$200~pc), and correlated it with star-formation activity (H$\alpha$).
They also found high $R_{21}$ in the center and bar-ends, in comparison to the bar and spiral arms.

\subsubsection{NGC~4254}

NGC~4254 is an interacting unbarred spiral galaxy (SAc) in the Virgo cluster. 
The angular resolution of 2.5$\arcsec$ is 168 pc. 
From Figure \ref{fig:co10co21_set2}, the molecular gas distribution appears asymmetric \citep[e.g.,][]{Hidaka:2002}. 
The center is bright and spiral arms are apparent in both transitions of CO, but the structures of the spiral arm and tightness differ between the northern and southern parts. 
The southern spiral arm extends beyond the coverage in CO (e.g., Figure \ref{fig:spitzer}). 

$R_{21}$ fluctuates between low and high-values along the spiral arms in the disk. 
Low-ratio and high-ratio gas encompass 46\% and 49\% of the area (blue histogram). 
Low-ratio and high-ratio gas include 44\% and 54\% of the CO(1-0) intensity distribution (green histogram), respectively.
Over the whole area, the mean $R_{21}$ is 0.72 while the intensity-weighted mean is 0.71.
From the histogram, the central kiloparsec ($R<0.5$ kpc) contains mostly high-ratio gas. 

\subsubsection{NGC~4321 \label{sec:ngc4321}}

NGC~4321 is a barred spiral galaxy (SABbc) in the Virgo cluster. 
The 2.5$\arcsec$ resolution corresponds to 173 pc. 
From Figure \ref{fig:co10co21_set2}, the molecular gas distribution shows a bright center and two spiral arms in both CO transitions. 
The spiral arms extend beyond the CO coverage (e.g., Figure \ref{fig:spitzer}). 

We note that for NGC~4321, the sensitivity in the CO(2-1) data between the northern and southern parts of the galaxy differs by about a factor of two (Figure \ref{fig:senrm}).
However, the analysis of the center, bar, and bar ends are all in the southern part where the sensitivity is better. 
Thus, we discuss the structural variations of $R_{21}$ in this galaxy as we do in the other galaxies.

$R_{21}$ is high in the center, lowers in the bar region, but fluctuates between low and high-values along the spiral arms (Figure \ref{fig:r21rgb_set5}). 
The southern spiral arm may be showing a potential transition from low to high-ratio gas along the gas flow across the spiral arm (where it runs horizontally in the image), assuming that the gas passes through the arm in the counter-clockwise direction.
Low-ratio gas encompasses 56\% of the disk while high-ratio gas makes up 35\% by area (blue histogram). 
Likewise, low-ratio gas contains 49\% of the CO(1-0) intensity and high-ratio gas contains 44\% (green histogram). 
Most of the molecular gas distribution in NGC~4321 thus has low-ratios by area and by CO(1-0) intensity. 
Over the whole area, the mean $R_{21}$ is 0.70 while the intensity-weighted mean is 0.71.
The central 1-2~kpc is mostly high ratio gas.

\subsubsection{NGC~4536 \label{sec:ngc4536}}

NGC~4536 is a barred spiral galaxy (SABbc). 
At its distance, the angular resolution of 2.5$\arcsec$ corresponds to a spatial resolution of 176 pc. 
From Figure \ref{fig:co10co21_set2}, the center is bright in both CO transitions and has two spiral arms which extend beyond the CO coverage (e.g., Figure \ref{fig:spitzer}). 
The bar is likely running nearly along the minor axis of the disk \citep{Diaz-Garcia:2016}, but because the bar appears substantially shorter due to this projection, its orientation can be debated \citep{Mazzalay:2014}. 

$R_{21}$ is high in the center but fluctuates between low and high-values along the spiral arms (Figure \ref{fig:r21rgb_set5}). 
The peak of the $R_{21}$ distribution by number is in low-ratio gas (blue histogram), but the peak shifts to high-ratio gas when weighted by CO(1-0) intensity (green histogram).
Low-ratio gas and high-ratio gas encompass 68\% and 28\% of the area. However, low-ratio gas contains 31\% of the CO(1-0) intensity distribution, with high-ratio gas containing 53\%. 
Over the whole area, the mean $R_{21}$ is 0.62 while the intensity-weighted mean is 0.79.
From the histograms, there is molecular gas concentrated in the central region, with 
the central 1-2~kpc containing predominantly high/very-high ratio gas.

\subsubsection{NGC~4579}

NGC~4579 is a barred spiral galaxy (SABb). 
At its distance, the angular resolution of 2.5$\arcsec$ corresponds to a spatial resolution of 199 pc. It is the farthest galaxy in the sample. 
From Figure \ref{fig:co10co21_set2}, the central region is bright in both CO transitions and is surrounded by spiral arms.
There is a concentration of gas in the central region. 
Its structure is spatially resolved and in the CO(1-0) map, it also looks like the roots of offset ridges. 
We treat it as a central concentration but since the transition from a central concentration to offset ridges is continuous, the boundary cannot be defined clearly. 

$R_{21}$ is high in the center.
The central gas concentration and high ratio region is also larger than the nominal definition of the central 1~kpc.
Indeed, the histogram for the central 2~kpc explains most of the HRG.
$R_{21}$ appears to fluctuate along the spiral structure (Figure \ref{fig:r21rgb_set6}). 
The distribution of the $R_{21}$ in the histogram weighted by CO(1-0) intensity appears doubly-peaked, signifying the difference in the $R_{21}$ distributions between the central region and spiral arms structure.
The peaks of the distributions by number and when weighted by CO(1-0) are both in low-ratio gas. 
Low-ratio gas and high-ratio gas encompass 88\% and 10\% of the area, respectively (blue histogram). 
However low-ratio gas contains 74\% of the CO(1-0) intensity while high-ratio gas contains 17\%. 
Over the whole area, the mean $R_{21}$ is 0.54 while the intensity-weighted mean is 0.63.
From the histograms, the central 1-2~kpc contains mostly high/very-high ratio gas, while the spiral arms contain low-ratio gas.

\subsubsection{NGC~4826}
NGC~4826 is an unbarred spiral galaxy (SAab). 
At its distance, the angular resolution of 2.5$\arcsec$ corresponds to a spatial resolution of 53 pc. It is the closest galaxy in the sample. 
The CO coverage of this galaxy in linear size is around the size of the central regions of the other galaxies in the sample. 
The molecular gas properties presented thus may be more comparable to the centers of galaxies. 
From Figure \ref{fig:co10co21_set2}, 
this central region is prominent and bright in both CO transitions. 

$R_{21}$ is high in the center but fluctuates between low and high in the surrounding regions. 
Low-ratio and high-ratio gas comprise similar portions of the galaxy by area, at 47\% and 49\% respectively.
However, low-ratio gas contains 36\% of the CO(1-0) intensity, while high-ratio gas contains 56\%. 
Over the whole area, the mean $R_{21}$ is 0.72 while the intensity-weighted mean is 0.76.
Given the small linear extent of the molecular gas, this skewed distribution towards high-ratio gas may be consistent with those in the central regions of the other galaxies, especially in the barred galaxies (see Figure \ref{fig:R21_size}).
This might indicate that once the gas is concentrated in the central regions, $R_{21}$ becomes high independent of the cause of the concentration.

\begin{deluxetable*}{cllcccc}
\tablecaption{CO(2-1) and CO(1-0) 12m+07m+TP data cube parameters before and after smoothing to 2.5$\arcsec$ angular resolution. The listed RMS noises are measured in high-sensitivity regions of emission-free channels.
\label{tab:cubeparam}}
\tablehead{
\colhead{(1)} & \colhead{(2)} & \colhead{(3)} & \colhead{} & \colhead{(4)} & \colhead{(5)} & \colhead{(6)}\\ 
\colhead{Galaxy} & \colhead{{FoV}} & \colhead{Beam Size} & \colhead{} & \multicolumn{3}{c}{RMS Noise (1$\sigma$)} \\
\cline{5-7}
\colhead{} & \colhead{Map Area, P.A.} & \colhead{b$_{\rm{maj}}$, b$_{\rm{min}}$, P.A.} & \colhead{} & \colhead{Native Res.}  & \multicolumn{2}{c}{Smoothed (2.5$\arcsec$)}  \\
\colhead{} & \colhead{[$\arcmin \times \arcmin$, $\arcdeg$]} & \colhead{[$\arcsec$, $\arcsec$, $\arcdeg$]} & \colhead{} & \colhead{[mK]} & \colhead{[mK]} & \colhead{[M$_\odot$ pc$^{-2}$]}}  
\startdata
\multicolumn{7}{c}{CO(2-1) Data Cube Parameters} \\
\cline{1-7}
 NGC~0628 & 4.3 $\times$ 2.9, 135 & 1.48, 0.88, -61.04  &&74.9  & 24.9 & 0.8, 1.8 \\
 NGC~1097 & 4.4 $\times$ 2.5, 145 & 1.54, 1.17, -85.48  &&  46.2  & 18.5 & 0.6, 1.3  \\ 
 NGC~1512 & 3.1 $\times$ 2.2, 37& 1.23, 0.99,  ~85.14  &&  71.9  & 21.8 & 0.7, 1.6  \\ 
 NGC~3351 & 2.6 $\times$ 2.5, 0 & 1.44, 1.20, -78.83  &&  92.2 & 36.2 & 1.1, 2.6  \\ 
 NGC~3521 & 4.9 $\times$ 2.1, 160 & 1.39, 1.12, -81.01 &&  37.9  & 13.6 & 0.4, 1.0  \\ 
 NGC~3621 & 4.2 $\times$ 2.3, 160  & 1.55, 1.13, -79.09  &&  31.8  & 12.2 & 0.4, 0.9  \\ 
 NGC~3627 & 4.2 $\times$ 2.3, 0 & 1.16, 1.00,  ~44.23  &&  92.0 & 28.1 & 0.9, 2.0  \\ 
 NGC~4254 & 3.5 $\times$ 3.0, 0 & 1.64, 1.17,  ~{5}7.54  &&  53.1  & 22.4 & 0.7, 1.6  \\ 
 NGC~4321 & 3.6 $\times$ 3.1, 90 & 1.13, 1.00, -30.25  &&  102.5 & 28.5 & 0.9, 2.1  \\ 
 NGC~4536 & 4.2 $\times$ 1.8, 119 & 1.55, 1.22, -71.86  &&  25.3  & 10.0 & 0.3, 0.7  \\ 
 NGC~4579 & 3.0 $\times$ 2.2, 52 &1.52, 1.33, -73.31   &&  39.3 & 16.8 & 0.5, 1.2  \\ 
 NGC~4826 & 2.8 $\times$ 1.8, 112 &1.35, 1.18, -23.85  &&  53.6 & 19.1 & 0.6, 1.4  \\ 
\cline{1-7}
\multicolumn{7}{c}{CO(1-0) Data Cube Parameters} \\
\cline{1-7}
 NGC~0628 & 4.3 $\times$ 3.3, 135 &2.42,  1.94,  -56.77  && 63.8 & 49.0 & 1.1  \\
 NGC~1097 & 4.8 $\times$ 2.8, 145 &2.11,  1.69,   89.44 && 75.7 & 47.4 & 1.0  \\
 NGC~1512 & 3.1 $\times$ 2.5, {37} &2.24,  1.70,   88.36 && 60.6 & 39.9 & 0.9 \\
 NGC~3351 & 2.9 $\times$ 2.8, 0 &2.10,  2.02,  -72.57 && 60.0 & 42.0 & 0.9  \\
 NGC~3521 & 5.3 $\times$ 2.5, 160 &2.18,  1.95,   88.84 && 67.1 & 47.5 & 1.0  \\
 NGC~3621 & 4.4 $\times$ 2.9, 160 &2.16,  1.64,   84.68  && 75.5 & 47.0 & 1.0  \\
 NGC~3627 & 4.4 $\times$ 2.9, 0 &2.17,  2.08,  -46.96 && 68.8 & 50.4 & 1.1  \\
 NGC~4254 & 4.0 $\times$ 3.7, 0 &2.41,  1.95,  -49.85 && 68.9 & 52.8 & 1.1 \\
 NGC~4321 & 4.0 $\times$ 3.6, 90 &2.17,  1.95,  -58.89 && 68.7 & 48.2 & 1.0 \\
 NGC~4536 & 4.5 $\times$ 2.2, 119 &2.10,  2.01,   85.63 && 59.2 & 41.3 & 0.9\\
 NGC~4579 & 3.1 $\times$ 2.6, 52 &2.20,  1.93,  -51.86 && 74.0 & 52.2 & 1.1 \\
 NGC~4826 & 3.1 $\times$ 2.1, 112 &2.45,  2.02,  -22.37 && 55.1 & 44.4 & 1.0
\enddata
\tablecomments{
(1) Galaxy (2) Rectangular field-of-view (FoV) enclosing the region out to where the sensitivity drops to 50\% of the peak sensitivity (40\% for NGC~4321 CO(2-1)).  The position angle (PA) is along the major axis of the map, counterclockwise from North.
(3) Beam Size.
(4) RMS Noise for the 12m+07m+TP data in 5 km/s channel at the native resolution.
(5)(6) Same as (4) but at the smoothed 2.5$\arcsec$ resolution and converted to mass surface density assuming $\alpha_{\rm CO(1-0)}$ = 4.35 and fiducial $R_{21}$=0.7 and $R_{21}$=0.3, respectively. The sensitivity limit becomes more important for lower brightness regions, which tend to have lower ratio.}
\end{deluxetable*}

\begin{deluxetable*}{ccccccccccc}[h]
\tablecaption{Area and intensity-weighted $R_{21}$, and the molecular gas surface density within and outside of the fiducial 1~kpc central region.}
\tablehead{
\colhead{(1)} & \colhead{(2)} & \colhead{(3)} & \colhead{(4)} & \colhead{(5)} & \colhead{(6)} & \colhead{(7)} & \colhead{} & \colhead{(8)} & \colhead{(9)} & \colhead{(10)} \\
\colhead{Galaxy} & \colhead{Ridge} & \colhead{$\Sigma^{\rm{center}}_{\rm H_2}$} & \colhead{$\Sigma^{\rm{outside}}_{\rm H_2}$} & \multicolumn{3}{c}{$R_{21}$ by Area} & \colhead{} & \multicolumn{3}{c}{$R_{21}$ by $I_{\rm CO}$} \\ 
\cline{5-7} \cline{9-11} 
 \colhead{} & \colhead{} &  \colhead{[M$_{\odot}$ pc$^{-2}$]} & \colhead{[M$_{\odot}$ pc$^{-2}$]} & \colhead{Mean} & \colhead{LRG [\%]} &  \colhead{HRG [\%]} & \colhead{} & \colhead{Mean} & \colhead{LRG [\%]} & \colhead{HRG [\%]}}
\startdata
\hline
NGC0628 & --         & 40           & 36 & 0.65 & 67 & 26 && 0.60 & 77 & 19\\
NGC1097 & L & 647          & 75 & 0.63 & 70 & 26 && 0.71 & 47 & 50\\
NGC1512 & S & 61           & 13 & 0.72 & 50 & 42 && 0.79 & 31 & 61\\
NGC3351 & S & 319          & 22 & 0.69 & 58 & 38 && 0.77 & 36 &62\\
NGC3521 & --         & 24           & 41 & 0.69          & 52 & 45          && 0.70          & 50          & 49\\
NGC3621 & --         & 21           & 17 & 0.76 & 35 & 58 && 0.78          & 26 & 67\\
NGC3627 & L & 394 & 41 & 0.70 & 57 & 38 && 0.71          & 48 & 50\\
NGC4254 & --         & 266          & 70 & 0.72 & 46 & 49 && 0.71          & 44 & 54\\
NGC4321 & L & 490          & 77 & 0.70 & 56 & 35 && 0.71 & 49 & 44\\
NGC4536 & S & 403          & 20 & 0.62 & 68 & 28 && 0.79          & 31 & 53\\
NGC4579 & S & 240          & 31 & 0.54 & 88          & 10          && 0.63 & 74          & 17\\
NGC4826 & --         & 200          & 62 & 0.72 & 47 & 49          && 0.76          & 36          & 56\\
\enddata
\tablecomments{\\ 
(1) Galaxy. \\ 
(2) Long or short (``L" or ``S") offset ridges for barred galaxies, based on the molecular gas morphology (Section \ref{sec:morphclass}, Figure \ref{fig:R21_arrangedbygasmorph}).``-" stands for unbarred galaxies. \\ 
(3) Average molecular gas surface density within the central 1~kpc using $\alpha_{\rm CO}$=4.35 in each galaxy. \\
(4) Same as (2) but outside of the central 1~kpc. \\ 
(5-7) Mean $R_{21}$, and percentage of molecular gas in LRG and HRG across the whole area of the galaxy.  \\
(8-10) Same as (5-7) but weighted by $I_{\rm CO(1-0)}$.}
\label{tab:percentages}
\end{deluxetable*}

\section{Discussion \label{sec:disc}}
This work presents high-resolution $R_{21}$ maps of a sample of nearby spiral galaxies, an early result from the ALMA-FACTS survey.
While we plan on more quantitative analyses in our subsequent papers, the $R_{21}$ maps already offer the first opportunity to compare the variations of $R_{21}$ amongst galaxies at high-resolution ($\sim$ 100-200~pc). 
The $R_{21}$ maps appear to show some general trends with respect to galactic morphology.
While, in general, discussions on morphology tend to be empirical, we attempt to characterize the general trends in terms of galactic structures.
In particular, the galactic centers, bars, and spiral arms appear to play their own distinct roles in the organization of the molecular gas conditions traced by $R_{21}$, in addition to their mass distributions as often discussed in the literature.
These empirical characterizations also seem consistent with the results in recent case studies at high-resolution ($\sim$ 100-200~pc) \citep[e.g.,][]{den-Brok:2021, Egusa:2022, Maeda:2022, den-Brok:2023}

We will first explain an idea for the stellar and gas morphology-based classifications and supporting dynamical models in the literature (section \ref{sec:morphclass}).
We then discuss the behaviors of $R_{21}$ in the galactic centers, bars, bar ends, and spiral arms, as well as radial and azimuthal variations of $R_{21}$ along this empirical classification (section \ref{sec:struct}).

\subsection{Morphology-Based Classification\label{sec:morphclass}}

For characterizing general trends in $R_{21}$, we rearrange the sample galaxies based on (1) their optical morphologies and (2) molecular gas morphologies (i.e., molecular gas distributions) with emphasis on bar and spiral arm structures.
Their optical morphologies are classified as barred (SB and SAB) and unbarred spiral galaxies (SA), which appear to primarily determine the behaviors of $R_{21}$ across the disks.
For barred galaxies, there is a suggested picture on the gas evolutionary sequence based on hydrodynamical simulations and observations \citep{Sakamoto:1999, Sheth:2005, Yu:2022}, which appears to help understand the behaviors of $R_{21}$ amongst the barred galaxies.  
We summarize this sequence and discuss our arrangement of the sample galaxies.

\subsubsection{Bar-Driven Gas Transport \label{sec:gasevol}}

In the bar-driven gas transport model, the stellar bar potential organizes the gas motions to form the offset ridges on the leading side of the bar. 
The interactions between the gas along the offset ridges (e.g., shocks) and/or with the stellar potential (e.g., torque) cause the gas to lose angular momentum, and the gas falls towards the center from the offset ridges \citep{Wada:1994, Sheth:2002}.
This picture is supported by observations.
\citet{Sakamoto:1999} and \citet{Sheth:2005} studied the central regions of barred spiral galaxies, and found that the central kiloparsec of barred spiral galaxies have higher concentrations of molecular gas than their unbarred counterparts. 
They, and others in the literature, attributed the concentrations of molecular gas in their central regions to the bar-driven gas transport by stellar bars to the centers \citep[][]{Sakamoto:1999, Sheth:2005, Garcia-Burillo:2009, Sormani:2023}.
In this process, the amount of gas in the bar should decrease, if no gas is inflowing from outside the bar.

\subsubsection{Classification \label{sec:class}}
\begin{figure*}[h]
    \centering
    \rotatebox{90}{
    \begin{minipage}{\textheight}
    \centering
    \includegraphics[width=1\textwidth]
    {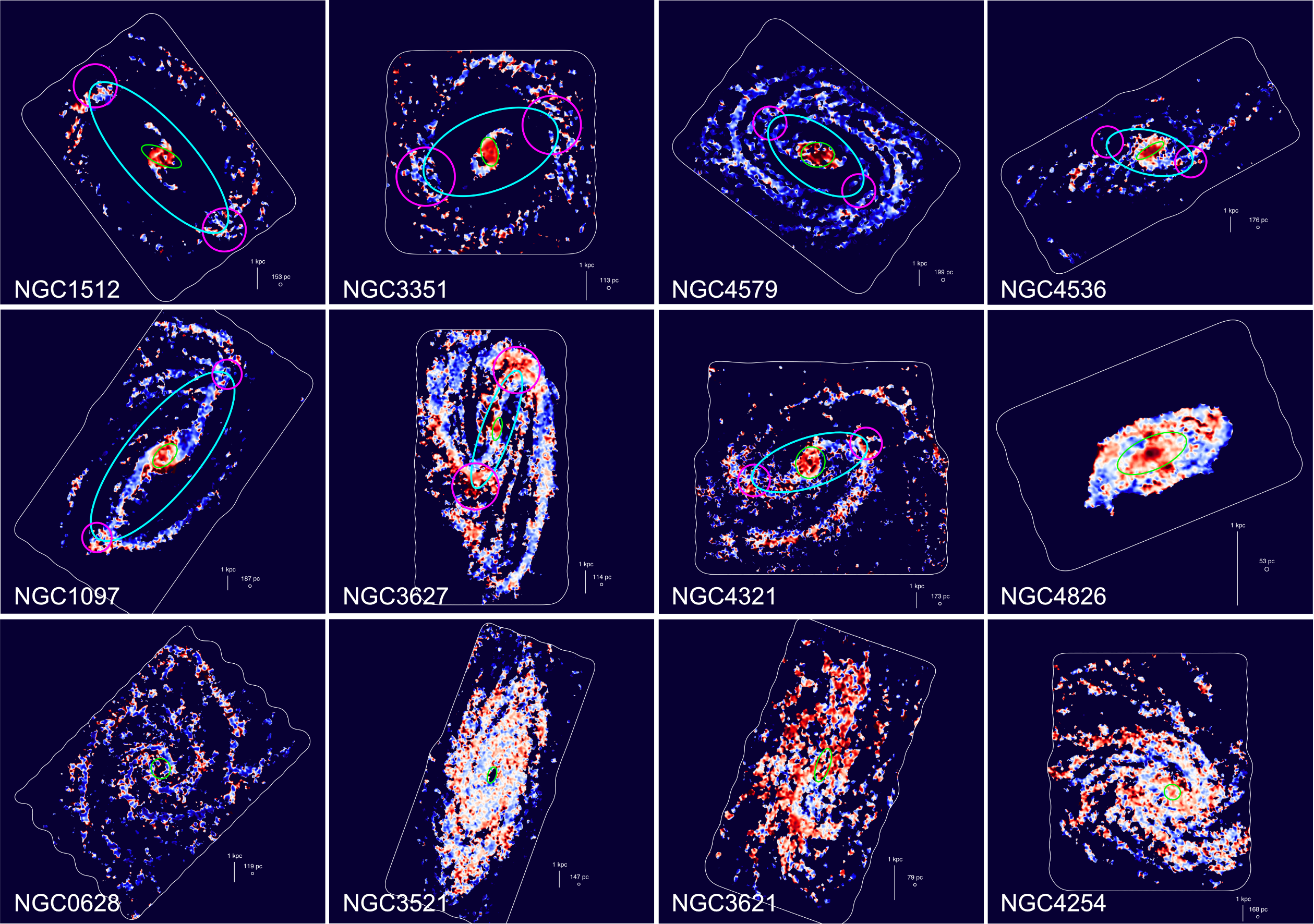}
    \caption{The $R_{21}$ maps arranged according to (1) optical morphology (SB, SAB, SA) and (2) gas morphology for the barred galaxies, following the gas morphological sequence presented in section \ref{sec:gasevol}.
    The adopted region definitions for the galactic structures are overlaid (section \ref{sec:r21_structures}). 
    The central regions (1 or 2~kpc) and the bar ends (2~kpc circles) are enclosed in green and magenta, respectively. 
    The bar regions are enclosed in the cyan ellipses, but do not include the bar ends.}
    \label{fig:R21_arrangedbygasmorph}
    \end{minipage}
    }
\end{figure*}

Figure \ref{fig:R21_arrangedbygasmorph} shows the $R_{21}$ maps from Section \ref{sec:results} arranged primarily by their optical morphological classifications (SB, SAB, or SA), and secondarily by their gas morphology for the barred galaxies.
The secondary part is based on the presence or absence of gas along the offset ridges. 
Even when the offset ridges are not present, all of the barred galaxies have central condensations of gas, and thus fit within the bar-driven gas transport picture, if the absence of the ridges is due to the gas inflow.
Hence, using the optical and gas morphological classifications, we arrange the $R_{21}$ maps (ordered by galaxy number) starting from barred galaxies with shorter offset ridges, i.e., lacking molecular gas between the central region and spiral arms, to barred galaxies with longer offset ridges and having clearer spiral arms extending from the bar ends, and then to the unbarred galaxies.

\begin{figure*}[h]
    \centering
    \includegraphics[width=1\textwidth]{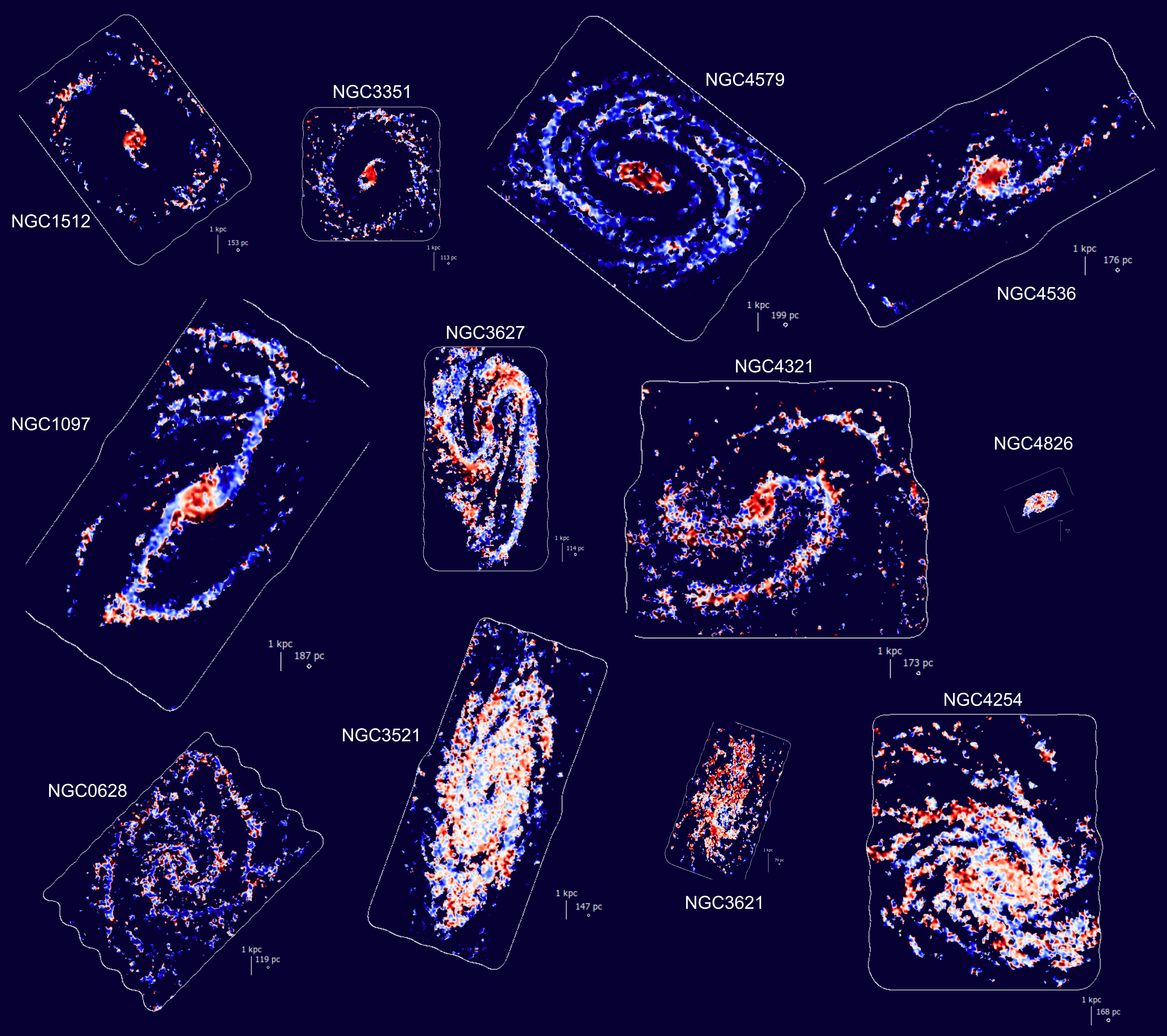}
    \caption{Same as Figure \ref{fig:R21_arrangedbygasmorph} but showing the relative physical sizes of the galaxies.}
    \label{fig:R21_size}
\end{figure*}

In the arranged gas morphological sequence of Figure \ref{fig:R21_arrangedbygasmorph}, the barred galaxies NGC~1512, NGC~3351, NGC4579, and NGC~4536 (top row) have shorter offset ridges and/or show an absence of gas between the central regions and their surrounding spiral arm structures.
Following these, the barred galaxies NGC~1097, NGC~3627, and NGC~4321 (left middle row) show longer offset ridges and have clearer spiral arms extending out of the bar ends. 
The rest of the galaxies are unbarred (the bottom row and the rightmost in the middle row).
Figure \ref{fig:R21_size} is the same as Figure \ref{fig:R21_arrangedbygasmorph}, but shows the galaxies on the same physical scale.
NGC~4826 (unbarred) is placed succeeding the barred galaxies (middle row) since it has a central gas condensation/region with a linear size more comparable to those of the barred galaxies (e.g., Section \ref{sec:r21} and Figure \ref{fig:R21_size}).

The barred galaxies have higher average $I_{\rm CO(1-0)}$ surface brightness in their central regions than the unbarred galaxies. 
If we simply assume the Milky-Way CO-to-H$_{2}$ conversion factor of $\alpha_{\rm CO}$ = 4.35 M$_{\odot}$pc$^{-2}$ $\cdot [\rm K\cdot km \cdot s^{-1}]^{-1}$ \citep{Bolatto:2013}, then this translates to higher average molecular gas surface densities in the centers of the barred galaxies (Table \ref{tab:percentages}).
This is consistent with \citet{Sakamoto:1999} and \citet{Sheth:2005}, which was interpreted as an outcome of bar-driven gas transport towards the center.
We note the conversion of surface brightness to molecular mass has the caveat that $\alpha_{\rm CO}$ may vary especially in the galaxy centers \citep[e.g.,][]{Sandstrom:2013, Teng:2023}.

In the following sections, we discuss $R_{21}$ along this sequence to understand the observed trends in $R_{21}$ as a function of structure.

\subsection{Structural Dependence of $R_{21}$ \label{sec:struct}}
We discuss the overall trends in $R_{21}$ seen amongst the barred and unbarred galaxies, and then within each galactic structure (center, bar, bar ends, and spiral arms). 
Figure \ref{fig:radialdist_arrangedbygasmorph} shows the radial distributions in two ways: as scatter plots to show the sampling (i.e., data density; top) and as density plots with the same data normalized by the peak number of points per radial bin to show the radial trend (bottom). 
They are arranged in the same order as Figure \ref{fig:R21_arrangedbygasmorph}.
The binned means (area-weighted) are shown in blue and orange for the barred galaxies with short or long ridges respectively, and teal for unbarred galaxies. The secondary x-axis is in arcseconds, and the blue dotted line indicates the bar length.

Overall, the barred galaxies (top and left middle rows) follow a general trend: $R_{21}$ is high in the center ($R_{21} \sim$ 1.0), low along the bar ($R_{21} \sim 0.6-0.7)$, increases at the bar ends ($R_{21} \gtrsim$ 0.7), and then lowers beyond the bar ends or flattens, with fluctuations in the spiral arms. 
The higher ratios in the bar ends are remarkable, given that the bar end areas are only in a small azimuthal range but still appear as high $R_{21}$ in these azimuthally-averaged plots.
The radial distributions of $R_{21}$ in the unbarred galaxies (right middle and bottom rows) look similar to the outer parts of the barred galaxies \citep{Komugi:2025}. 

There are also some variations in the azimuthal direction in the sample galaxies. 
However, due to the sensitivity limitations, most interarm regions of the galaxies are not readily detected (Figure \ref{fig:R21_arrangedbygasmorph}). 
Variations in the azimuthal direction can be seen as scatters at each radius in the radial distributions. 
With these trends in mind, we discuss $R_{21}$ as a function of structure in the following sections.

\subsubsection{Central Regions: High $R_{21}$ \label{sec:center}}

It has been known that barred galaxies often have high gas concentrations in their central regions \citep{Sakamoto:1999, Sheth:2005, Yu:2022}.
The same central gas concentrations are apparent in the barred galaxies in our observations (Figures \ref{fig:co10co21_set1}-\ref{fig:co10co21_set2}).
In addition, the sample of the unbarred spiral galaxies do not show such gas concentrations at the centers besides NGC~4826. 
This study (section \ref{sec:r21}) shows that the central 1-2 kiloparsec regions within barred galaxies also comprise of high/very-high ratio gas, and that the gas physical conditions change within the central gas condensations. 
$R_{21}$ is elevated significantly and consistently in the centers, which has also been found in earlier studies, e.g. \citet{Braine:1992, Leroy:2009, den-Brok:2021}.
The elevated line-ratios in their galactic centers are clear in the radial distributions.

The presence of high-ratio gas across these central regions indicates that the molecular gas there has higher $n_{\rm{H2}}$ and $T_{\rm{k}}$ and is thus denser and/or warmer than the typical molecular gas conditions, ($n_{\rm H_2}$, $T_{\rm k}$)=(300~cm$^{-3}$, 10~K), in the disks \citep{Koda:2012}. 

The high ratios may be the result of dynamical processes and/or on-going star formation in the galaxy centers. 
As a dynamical process, the high ratios in the central regions of the barred galaxies across the sample could be due to the bar-driven gas transport to the centers, discussed in section \ref{sec:gasevol}.
The influence of the bar may compress the gas (increase in gas density) and/or trigger star formation in galactic centers (increase in gas kinetic temperature by stellar feedback) and lead to the high-ratios. 
The enhancement of star formation in the centers of barred galaxies has been studied e.g., in \citet[][]{Ho:1997}.

Amongst the sample galaxies, NGC~1097, NGC~3627, NGC~4579, and NGC~4826 (as well as NGC~3621) are classified as having AGN \citep{Veron-Cetty:2010} and this may also be a cause of the high $R_{21}$ there. 
However, even the galaxies without AGN often show high $R_{21}$ in the central $\sim 1-2$kpc when they have the central gas concentrations. 
Thus, AGN may not be the only cause of high $R_{21}$ in central regions.

It may also be worth noting that NGC~4826 is an unbarred spiral galaxy in the optical classification, but has a central concentration of high-ratio gas.
Although it is only one case, its presence may indicate that the high ratio is due to the local environment of the central region (e.g., strong shears or tidal forces), rather than the compression due to the infalling gas.

\subsubsection{Along the Bar: Low $R_{21}$  \label{sec:alongbar}} 
\begin{figure*}[h]
    \centering
    \includegraphics[width=0.95\textwidth]{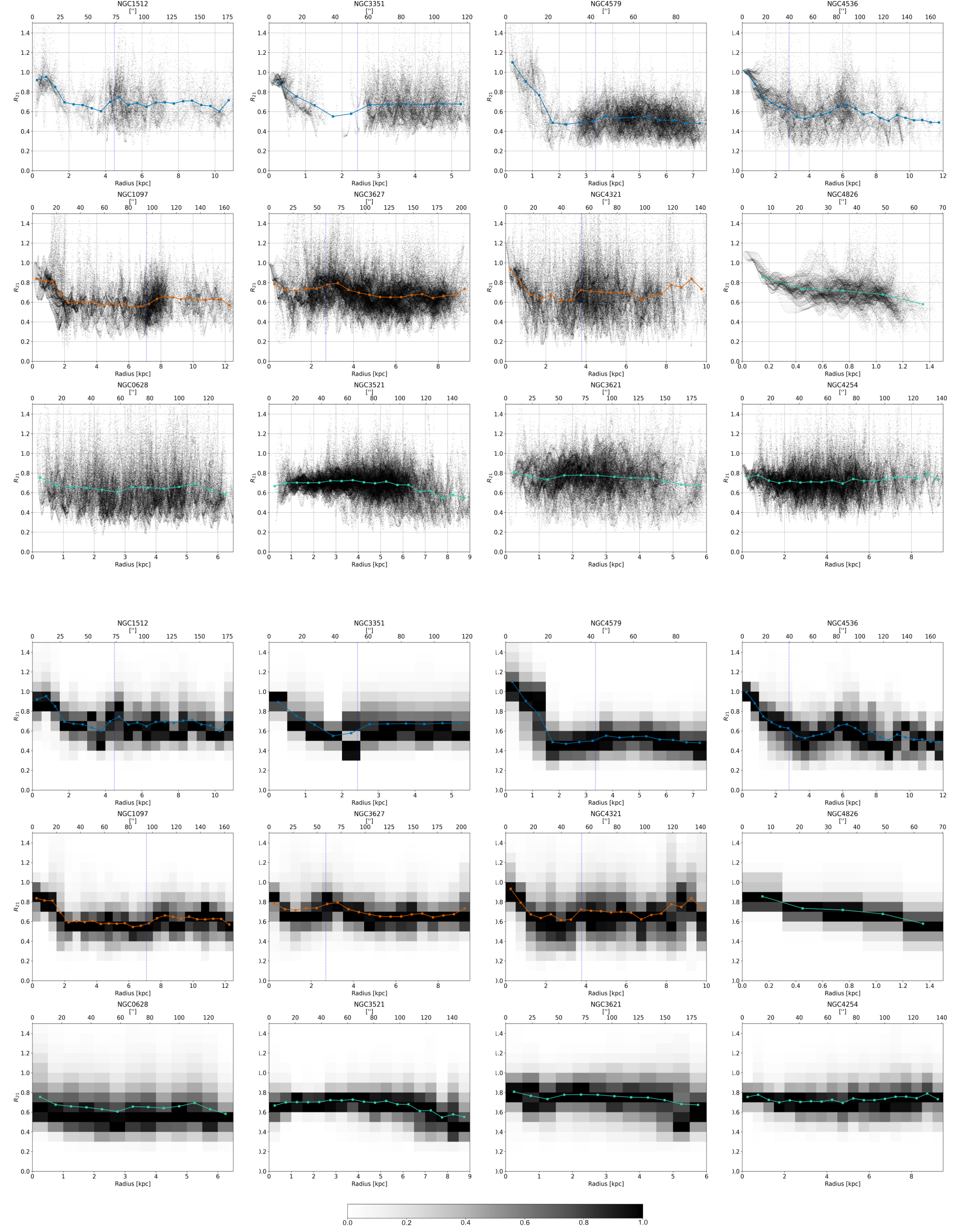}
    \caption{{{Radial distributions of $R_{21}$ arranged in the same order as Figure \ref{fig:R21_arrangedbygasmorph}, shown as  scatter plots (top) and with the same data normalized by the peak number of data points in each radial bin (bottom).}
    The binned means (area-weighted) are shown in blue and orange for the barred galaxies with short or long offset ridges, and in teal for the unbarred galaxies. 
    The secondary-axis is in arcseconds and the blue dotted lines indicates the adopted bar lengths ($R_{\rm bar}$).}}
    \label{fig:radialdist_arrangedbygasmorph}
\end{figure*}

Bar regions overall show lower $R_{21}$ ($R_{21} \sim 0.6$). 
This is evident in the spatial and radial distributions of $R_{21}$ (Figures \ref{fig:R21_arrangedbygasmorph}-\ref{fig:radialdist_arrangedbygasmorph}), even though there is molecular gas concentrated along the offset ridges and bar (Figures \ref{fig:co10co21_set1}-\ref{fig:co10co21_set2}). 
This seems to be a general trend amongst barred spiral galaxies, including the ones in other studies: NGC~1300, NGC~1365, NGC~2903, NGC~3627, and NGC~5236 (M83) \citep[][]{Maeda:2022, Egusa:2022, den-Brok:2023, Koda:2025}.
The low $R_{21}$ in the bar region is seen even when the offset ridges do not have much gas ({NGC~1512, NGC~3351, and NGC~4579; top row} in Figures \ref{fig:R21_arrangedbygasmorph}-\ref{fig:radialdist_arrangedbygasmorph}).

The lower $R_{21}$ suggests that $n_{\rm H_2}$ and $T_{\rm k}$ remain low in the bar \citep{Koda:2012}, even when the gas is concentrated along the offset ridges.
This may explain why star formation efficiency is often low in the bar regions of galaxies \citep[e.g., ][]{Downes:1996, Koda:2006, Momose:2010, Maeda:2023}.
Recently, \citet[][]{Rodriguez:2024} analyzed the 3.3 $\mu$m polycyclic aromatic hydrocarbon (PAH) emission in 19 nearby star-forming galaxies to identify star clusters in their early dust-embedded phases. 
They also found that, in general, there is an absence of PAH emitters in bars (including our FACTS sample: NGC~1512, NGC~3351, NGC~3627, and NGC~4321). 

The generally low $R_{21}$ along the offset ridges may also be important in considering gas dynamics within the bar. 
It has been suggested that the offset ridges are regions of strong shocks \citep{Athanassoula:1992a, Athanassoula:1992b, Sellwood:1993}, which should naturally increase $n_{\rm H_2}$ and $T_{\rm k}$, and hence, $R_{21}$.
Instead, the gas concentration along the offset ridges may be caused by the mere crowding of gas orbits with only occasional collisions \citep{Wada:1994, Koda:2006}, which could keep $R_{21}$ low in general.
In the case of orbit crowding, the gas falls towards the center by the loss of angular momentum due to the gravitational torque of the stellar bar potential, as the gas resides longer on the leading side of the bar \citep{Wada:1994}.
This gas infall process would take longer than the rapid gas infall by the strong shocks, which could make offset ridges live longer, and appear more frequently in a sample of barred galaxies. 
In the future, it may be interesting to check the frequency of the offset ridges (from long to short to none), in a larger sample.

\subsubsection{Bar Ends: High $R_{21}$ \label{sec:barend}}
Bar ends are the transition regions from the offset ridges in the bar to spiral arms outside the bar.
The gas often builds up around the bar ends \citep[e.g., NGC~1097 and NGC~3627 in Figures \ref{fig:R21_arrangedbygasmorph}-\ref{fig:radialdist_arrangedbygasmorph}, see also][]{Sheth:2002}.
Even when the offset ridges fade away at their further ends, the CO emission often starts to re-appear around the bar ends and extends along the spiral arms (e.g., NGC~1512, NGC~3351).
Star formation is likewise evident at the bar ends in the optical images e.g. for NGC~1097 and NGC~3627 (Figures \ref{fig:r21rgb_set1} and \ref{fig:r21rgb_set4}). 
\citet[][]{Rodriguez:2024} show there is a clustering of 3.3 $\mu$m PAH emitters at the bar ends of NGC~3627.

The bar ends also show higher $R_{21}$ spatially and in the radial distributions often $R_{21} \gtrsim 0.7$) (Figures \ref{fig:R21_arrangedbygasmorph}-\ref{fig:radialdist_arrangedbygasmorph}). 
This is most apparent in the galaxies with longer offset ridges (see also section \ref{sec:r21_structures}).
{The high ratios again suggest that the molecular gas located there has relatively higher $n_{\rm H_2}$ and/or $T_{\rm k}$ \citep{Koda:2012}}. 
When there is not much gas in the bar region (as in the short-offset ridge galaxies) it is difficult to assess. 
NGC~1512 shows an increase towards the bar ends in its radial profile, but this is unclear in NGC~3351 and NGC~4579. NGC~4536 is affected by projection effects (section \ref{sec:ngc4536}).

The stellar bar pattern should have a constant pattern speed to maintain its linear structure over time.
The bar end is often considered to be around the co-rotation radius where the orbital speed of gas and stars are synchronized with the bar pattern speed (it is at $\sim$80\% of the co-rotation radius according to \citet{Athanassoula:1992a, Athanassoula:1992b}).
Once the gas gets into the bar end, it will likely reside there for a long time.
This explains the gas concentrations around the bar ends.
The combination of the high gas concentration and long residential time may trigger local gravitational collapses, which may explain the high $R_{21}$ and star formation \citep[e.g.,][]{Maeda:2025}.
This is, of course, a hypothesis and needs to be verified by future observational and theoretical studies.

\subsubsection{Spiral Arms \label{sec:spiralarm}}
Spiral arms are also where the gas builds up.
$R_{21}$ fluctuates between low and high values along the molecular gas spiral arms
(Figures \ref{fig:R21_arrangedbygasmorph}-\ref{fig:radialdist_arrangedbygasmorph}) \citep[see also e.g.,][]{Druard:2014, Koda:2025}.
It is remarkable that the fluctuations appear similar between the spiral arms in the barred galaxies and those of the unbarred galaxies.
High-ratio gas in these regions sometimes appear to be associated with HII regions, and thus, star formation.
This has been found in Galactic and extragalactic studies \citep[e.g.,][]{Sakamoto:1994, Seta:1998, Koda:2012, Koda:2020, Egusa:2022, Koda:2025}. 

These spatial correlations provide an important benchmark for understanding the interplay between molecular gas evolution and star formation in spiral galaxies.
However, their causality needs to be discussed with more detailed comparisons with multi-wavelength data or with future observations.
In fact, the galactic potential, such as the spiral arm potential, can control gas dynamics/flows, which can compress the gas, increase $R_{21}$, and trigger star formation.
On the other hand, the galactic structures can help in assembling the gas to form stars, and their feedback can increase $R_{21}$.

In the analysis of the nearby barred spiral galaxy M83, which had higher resolution and sensitivity, \citet{Koda:2025} recently found that $R_{21}$ changes systematically from a low ratio in the interarm regions to high ratio in the bar and spiral arms even without star formation. 
They concluded that the large-scale galactic dynamics plays an important role in the evolution of molecular gas, in addition to the local impacts of stellar feedback.
In this sample, NGC~1097, NGC~3627, and NGC~4321 show more coherent molecular spiral arms, with some parts possibly showing a similar transition from low to high $R_{21}$.

\subsubsection{$R_{21}$ by Structure \label{sec:r21_structures}}
$R_{21}$ clearly varies within each galaxy. 
The dependence on galactic structures (center, bar, bar ends, and spiral arms) is apparent in Section \ref{sec:r21}. 
For a more quantitative comparison, we define regions to enclose the structures in each galaxy (Figure \ref{fig:R21_arrangedbygasmorph}). 
We created masks for the center, bar, and bar ends.
Our adopted region definitions are not meant to be exact boundaries, but are an attempt to capture the rough areas of the structures of interest. 
We describe the definitions and $R_{21}$ distribution in each region below. 

For the central regions, we adopt either 1 or 2~kpc diameter areas around the galactic center (projected in the sky). 
We adopt the fiducial 1~kpc for all \emph{unbarred} galaxies. 
For the \emph{barred} galaxies, we showed that the size of the high-ratio region in the center sometimes extends beyond this in section \ref{sec:r21}. 
We adopt 2~kpc for all barred galaxies except for NGC3351 and NGC3627. 

For the bar and bar ends, we define regions based on the cataloged bar structure (cyan ellipse) in \citet{Herrera-Endoqui:2015}, which is based on Spitzer 3.6 $\mu$m from S4G \citep{Sheth:2010, S4G}. 
We use their on-sky PA$_{\rm bar}$, semi-major axis $R_{\rm bar}$, and ellipticity (see Table \ref{tab:sample}). 
The bar region is enclosed by the cyan ellipse but does not include the bar ends (magenta).
The bar ends are 2~kpc circles centered on the ``tips” of the ellipse.

We adopt the definitions of the spiral arms from \citet{Querejeta:2021}.
The area outside the center, bar, bar end, and spiral arms is analyzed as interarm regions.

Using these definitions, we compared the $R_{21}$ distributions for adjacent structure pairs (center-bar, bar-bar end, and bar end-spiral arm) and the spiral arm-interarm regions by performing the Kolmogorov-Smirnov (KS) two-sample test
using the \emph{ks\_2samp} function in \emph{scipy}. 
The null hypothesis is that the two samples originate from the same parent distribution. 
We applied this test including only every fifth pixel (Nyquist sampling) (1) within each galaxy and (2) all galaxies combined.
We find p-values $<0.05$ (majority lower than $<10^{-6}$) meaning the null hypothesis can be rejected in all the structure pairs compared in all galaxies, except in two cases: bar-bar end in NGC~1512 (p-value $\sim$ 0.4) and bar end-spiral arm in NGC~4536 (p-value $\sim$ 0.3). 
In NGC~1512 there is not much gas in the bar, and in NGC~4536 the bar-end definition is ambiguous due to its projection (section \ref{sec:ngc4536}). 
{(Note: with full-beam sampling, the NGC1097 bar-bar end and NGC4321 bar end-spiral arm comparisons also show p-values $>$ 0.05. However, p-values are still $<0.05$ in most cases, and the overall discussion does not change.)}

\begin{figure*}[h]
    \centering
    \includegraphics[width=0.5\textwidth]
    {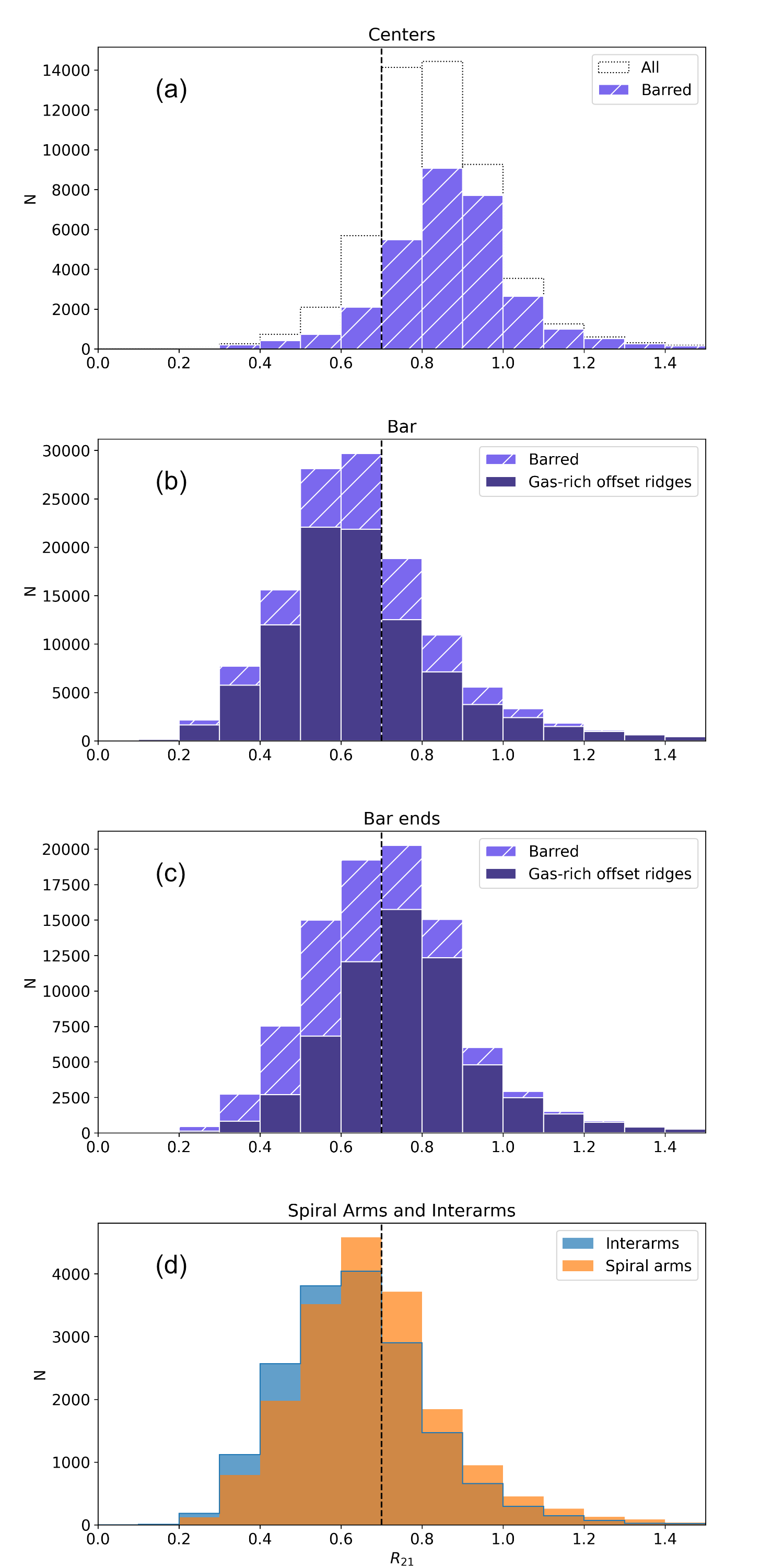}
    \caption{(a-d) $R_{21}$ distributions by number for the center, bar, bar ends, and spiral arms and interarm regions, respectively, using the definitions described in section \ref{sec:r21_structures} and Figure \ref{fig:R21_arrangedbygasmorph}. 
    The black dashed line is at $R_{21}$=0.7 for reference. 
    For (a-c) the histograms are shown for all galaxies (dotted), barred galaxies (purple, hatched), and those with the long offset ridges (dark purple) following the classification introduced in section \ref{sec:class}.}
    \label{fig:R21_structures_hist}
\end{figure*}

Figure \ref{fig:R21_structures_hist} shows the corresponding $R_{21}$ distributions.
They reflect the general trends discussed in section \ref{sec:struct}. 
Central regions of barred galaxies show high $R_{21} > 0.7$. 
Bars show mostly low $R_{21} < 0.7$. 
Bar ends for the galaxies with relatively longer offset ridges more often show high $R_{21} > 0.7$. 
We note that the longer offset ridged galaxies contribute the most in panels (c-d) because more emission is detected there (see section \ref{sec:barend}).

The KS test also suggests the arm-interarm distributions are different (p-values $< 10^{-6}$).
The histogram for the interarms is slightly but systematically shifted towards lower $R_{21}$ (Figure \ref{fig:R21_structures_hist}).
The contrast could be higher if the sensitivity is higher because low brightness/density gas, expected more in the interarms, tends to have lower $R_{21}$.
We note that our equal flux-based cut in CO(1-0) and CO(2-1) (SN$>$5) practically corresponds to a higher density threshold for CO(2-1) if $R_{21}$ is low (Table \ref{tab:cubeparam}, column (6)).

We note that \citet{Querejeta:2021} also defined the inner structures (the center, bar, and bar ends).
Their bar ends are defined as overlap regions of bar and spiral arms,
but are not identified when their bar is defined too short to reach the spiral arms.
This does not assure a consistent analysis across the galaxy sample, and thus, we adopted the above definitions.
The trends of $R_{21}$ discussed above for the centers and bars are the same when we use their definitions.

\section{Systematic Trends and Comparison to Literature \label{sec:comp2lit}}
In sections \ref{sec:results} and \ref{sec:disc}, we showed that $R_{21}$ systematically varies as a function of galactic structure in the twelve nearby spiral galaxies. 
Such structural variations were not previously clear from single-dish surveys \citep{den-Brok:2021, Yajima:2021, Leroy:2022}. 
\citet[][Paper II]{Komugi:2025} conducted the $R_{21}$ analysis using the ALMA-TP component only. 
They were not able to resolve the galactic structures with their coarser resolution, but found that most of the barred galaxies showed steeper radial gradients, whereas all the unbarred galaxies showed flat or shallower gradients.

This general trend of structure-dependent variations is consistent with the results from high-resolution ALMA case studies. 
For example, \citet{Egusa:2022} studied NGC~1365 at $\sim$ 200~pc resolution and found high $R_{21}$ $>$ 0.7 in the center while it is low $R_{21}$ $<$ 0.7 in the bar region. 
\citet{Maeda:2022} studied NGC~1300 at $\sim$100~pc resolution, finding $R_{21}$ to 
be low along the bar $\sim 0.50$, high $\sim 0.72$ in the bar end, and $\sim 0.60$ in the spiral arm.
While these studies covered only small regions within individual galaxies, the results are consistent with the general trend. 
Furthermore, \citet{den-Brok:2023} studied $R_{21}$ in the disk of NGC~3627 at $\sim$200~pc resolution and found a consistent trend, with high $R_{21}$ $\sim$ 0.90 in the center, low $\sim$ 0.70 along the bar, and then increasing again at the bar ends to $\sim$ 0.86, and high $\sim$ 0.75 in the spiral arms.
The same trend was also found in M83 at a higher 46~pc resolution \citep{Koda:2025}.
The high $R_{21}$ in the centers of the barred galaxies could explain the steeper radial gradients found by \citet{Komugi:2025}.

Most of these studies, including FACTS, have insufficient sensitivity to detect interarm emission; but when it is detected, it appears to have low-ratios.
The M83 study by \citet{Koda:2025} has higher sensitivity to detect faint interarm gas, which clearly exhibits low-ratios. 
As the trends amongst bright structures in M83 is similar to that of the FACTS galaxies and other high-resolution case-studies, we conclude that the observed trends in $R_{21}$ are common amongst all nearby (barred) star-forming spiral galaxies.

\section{Summary}

In this study, we analyzed the CO(2-1)/CO(1-0) line ratio in twelve nearby galaxies from the Fundamental CO(1-0) Transition Survey (FACTS) in conjunction with the PHANGS CO(2-1) survey.
We focused on empirical trends and showed that $R_{21}$ systematically varies as a function of galactic structures in twelve nearby galaxies. 
To characterize the observed structural variations, we made empirical classifications of the galaxies based on their optical morphologies (SB, SAB, or SA), and their molecular gas morphologies for the barred galaxies, depending on the presence or absence of molecular offset ridges. 
The molecular gas morphological sequence may be explained as an evolutionary sequence by the bar-driven gas transport model. We discussed $R_{21}$ in the context of this sequence.

From the radial distributions of $R_{21}$, the barred galaxies follow a general trend: $R_{21}$ is high in the central 1-2~kpc regions $\sim$ 1.0, low along the bar offset ridges $\sim$ 0.6, increases at the bar ends often $\gtrsim$ 0.7, and then declines in the rest of the disk with fluctuations along the spiral arms. 
The radial distributions of the unbarred galaxies fluctuate in $R_{21}$ and are similar to the outer parts (spiral arms) of the barred galaxies.

High $R_{21}$ in the central 1-2~kpc regions is always associated with the central gas condensations in the barred galaxies. NGC~4826 is the only unbarred spiral galaxy in our sample that shows the central condensation and high $R_{21}$.
Low $R_{21}$ along the bar offset ridges suggest that the gas density and/or temperature are lower there \citep{Koda:2012}, which may explain the reduced star formation efficiency along bars as suggested by previous studies.
Bar ends show high $R_{21}$, indicating increased density and/or temperature.
In spiral arms, $R_{21}$ fluctuates, presumably due to the impacts of star-formation activity or galactic dynamics.

The $R_{21}$ variations found in this study are consistent with previous high-resolution ALMA case studies (NGC 1300, NGC 1365, NGC 3627, and M83). 
In particular, while interarm emission is not well-detected in this study due to the sensitivity limitation, all the other structures show trends in $R_{21}$ similar to that observed in a recent case-study of M83 which had higher sensitivity and resolution \citep{Koda:2025}. 
The common trends in $R_{21}$ amongst the FACTS sample, M83, and a few other barred galaxies suggests that they may be common amongst all nearby (barred) star-forming galaxies.

We have shown that the variations in $R_{21}$ appear to depend on the galactic structures (the central regions, bar offset ridges, bar ends, and spiral arms).
The structure dependence may also suggest the importance of galactic dynamics on the molecular gas evolution and star formation.
We plan more detailed multi-wavelength comparisons in our future publications from the FACTS survey.

\clearpage

\bibliography{references.bib}
\bibliographystyle{aasjournal}

\clearpage
\begin{acknowledgments}
This work makes use of the following ALMA data: ADS/JAO.ALMA\#2012.1.00650.S, 2015.1.00925.S, 2015.1.00956.S, 2017.1.00886.L, and 2018.1.01651.S, 2022.1.00360.S.
ALMA is a partnership of ESO (representing its member states), NSF (USA) and NINS (Japan), together with NRC (Canada), MOST and ASIAA (Taiwan), and KASI (Republic of Korea), in cooperation with the Republic of Chile. The Joint ALMA Observatory is operated by ESO, AUI/NRAO and NAOJ.
The National Radio Astronomy Observatory is a facility of the National Science Foundation operated under cooperative agreement by Associated Universities, Inc..

This research has made use of the NASA/IPAC Infrared Science Archive, which is funded by the National Aeronautics and Space Administration and operated by the California Institute of Technology.
This research has made use of the NASA/IPAC Extragalactic Database (NED),
which is operated by the Jet Propulsion Laboratory, California Institute of Technology,
under contract with the National Aeronautics and Space Administration.

JK acknowledges support from NSF through grants AST-2006600 and AST-2406608.
F.M. is supported by JSPS KAKENHI grant Nos. JP23K13142 and JP23K20035.
S.K. is supported by JSPS KAKENHI grant No. JP25K07371.
\end{acknowledgments}

\end{document}